\documentclass[a4paper,fleqn,usenatbib]{mnras}

\usepackage{newtxtext,newtxmath}
\usepackage[T1]{fontenc}
\usepackage{ae,aecompl}

\usepackage{hyperref}

\usepackage{graphicx}	% Including figure files
\usepackage{amsmath}	% Advanced maths commands
\usepackage{amssymb}	% Extra maths symbols

\title[Gasoline 2]{Gasoline2: A Modern SPH Code}

\author[J. W. Wadsley et al.]{
James W. Wadsley,$^{1}$\thanks{E-mail: wadsley@mcmaster.ca}
Benjamin W. Keller,$^{1}$
and Thomas R. Quinn$^{2}$
\\
$^{1}$Department of Physics \& Astronomy, McMaster University, Hamilton, L8S 4M1, Canada\\
$^{2}$Astronomy Department, University of Washington, Seattle, Washington, USA\\
}

\date{Accepted XXX. Received YYY; in original form ZZZ}

\pubyear{2017}

\begin{document}
\label{firstpage}
\pagerange{\pageref{firstpage}--\pageref{lastpage}}
\maketitle

\begin{abstract}
  The methods in the {\sc Gasoline2} Smoothed Particle Hydrodynamics (SPH)
  code are described and tested.  
  {\sc Gasoline2} is the most recent version of the {\sc Gasoline}
  code for parallel hydrodynamics and gravity with identical 
  hydrodynamics to the {\sc Changa} code.
  As with other Modern SPH codes, we prevent sharp jumps in time steps,
  use upgraded kernels and larger neighbour numbers and employ local viscosity
  limiters.  Unique features in {\sc Gasoline2} include its
  Geometric-Density-Average Force expression, explicit Turbulent Diffusion
  terms and Gradient-Based shock detection to limit artificial viscosity.
  This last feature allows {\sc Gasoline2} to completely avoid
  artificial viscosity in non-shocking compressive flows.
  We present a suite of tests demonstrating the value of these features with
  the same code configuration and parameter choices used for production simulations.

\end{abstract}


\section{Introduction}
We present the {\sc Gasoline2}\footnote{{\sc Gasoline2} is a public code, and is
available for download at http://gasoline-code.com} smoothed particle
hydrodynamics (SPH) code; which is the culmination of improvements over several
years to the original {\sc Gasoline} SPH code \citep{Wadsley2004}.  The purpose
of this paper is to provide details of the modern SPH implementation that is
current in both {\sc Gasoline2} and the {\sc Changa} code \citep{Menon2015}.
{\sc Gasoline2} includes several unique features that substantially enhance the
accuracy of SPH on problems of astrophysical interest.
We explicitly note that several of these
features were introduced previously \cite[e.g.][]{Wadsley2008,Shen2010,Keller2014}.
The current work is the first to provide complete details and show
results on standard tests when all these features are combined.

SPH is a method for discretizing and solving Euler's equations; introduced by
two independent groups in \citet{Lucy1977} and \citet{Gingold1977}.   SPH
discretizes co-moving mass elements, making it a Lagrangian
method.   This has the advantage of making the tracing of fluid elements trivial.
This also makes SPH automatically concentrate resolution
elements where densities are largest, which is a common strategy for
adaptive codes in astrophysics.  
SPH has perfect Galilean
invariance and exact momentum, angular momementum and energy 
conservation due to its pairwise interactions.  In
order to handle shocks, SPH requires the use of artificial viscosity, typically
paired with a limiter \citep[e.g.][]{Balsara1995} to reduce the viscosity away
from shocks.
This makes SPH quite numerically robust compared to
Riemann-solver based codes.  SPH's particle nature is also ideal for
unification with N-body schemes for gravity \citep{Evrard1988,
  Hernquist1989}.
Lagrangian schemes benefit from time steps that depend only on local
wave speeds (e.g. sound speed) rather than absolute velocities making
them efficient in cold, self-gravitating systems (e.g. disks) as well as avoiding frame
and direction dependent behaviour.

SPH has become a pillar of numerical astrophysics, along with Eulerian
adaptive-mesh refinement (AMR) schemes based on Riemann solvers
such as ENZO \citep{Bryan1997}, FLASH
\citep{Fryxell2000}, and RAMSES \citep{Teyssier2002}.
% ENZO -- cite the first one to show it was developed first No-one is going to
% use our paper to find the most recent AMR code papers
Recently these methods have been joined by hybrid schemes that attempt to merge
the best features of both classes of method, including AREPO
\citep{Springel2010a} and GIZMO \citep{Hopkins2015} (see also: \citealt{Gaburov2011}).
%  Add ref to Garcia-senz??

While traditional SPH (as presented in \citet{Monaghan1992,Monaghan2005}, {\sc Gadget2}
\citep{Springel2002}, and the original {\sc Gasoline}) has many attractive
features, it became clear that it also had some significant shortcomings,
especially when dealing with multiphase fluids.
These problems were highlighted  
in \citet{Agertz2007} through the application of several major codes to a suit
of novel test problems.   
%There are quite a few ``limitations of SPH'' papers beyond agertz
Referring to SPH formalisms prior to $\sim$ 2008
as ``traditional'' is fairly common in the literature
\citep[e.g.][]{Gibson2009,Hopkins2013}.   Other common labels include
``standard'' SPH \citep{Springel2010b} and ``classic'' SPH \citep{Read2010,Bauer2012}.
For the purposes of this work, we shall refer to all SPH in common usage
prior to 2008 as ``traditional'' SPH.    We chose 2008 because it was the year
that artificially supressed mixing was first identified and then resolved for SPH
\citep{Wadsley2008,Price2008}, primarily in response to \citet{Agertz2007}.
Some of the issues raised were already known and SPH variants were proposed to
address them \citep[e.g.][]{Ritchie2001}, but were not widely adopted.   It should also be
noted that {\sc Gadget2} \citep{Springel2002} differed
from other contemporary SPH codes in employing a formulaton derived from a
Lagrangian (first proposed by  \citealt{Nelson1994}).  This requires iterations to
satisfy a constraint but offers specific benefits in simplifying entropy and
energy conservation (as discussed further in section~\ref{sec:method}).  This
change does not help resolve its primary shortcomings such as the mixing issue.
In what follows we present our own take on the state of SPH and important
developments, focussing on those that influenced the design of {\sc Gasoline2}.
We refer the reader to recent reviews of SPH for additional perspectives, such as \citet{Springel2010b,Price2012a,Rosswog2015}.

SPH and Eulerian codes generally agree well in simulations that
involve supersonic flow
\citep[e.g. comparison projects by][]{Kitsionas2009,Price2010,Bauer2012},
high-Mach number shocks \citep{Tasker2008,Sijacki2012},
and gravitational structure formation \citep{OShea2005}.
%Pilkington?????  
However, in problems involving multiphase gas in near pressure equilibrium,
SPH has been found to have a number of problems
\citep[e.g.][]{Agertz2007, Read2010}.  These problems manifest as
suppressed growth for fluid instabilities in shearing flows.
Work by many authors, as described below, has demonstrated
that these problems primarily arise from three sources: insufficient
mixing, surface tension at density discontinuities, and slow numerical
convergence.  The last issue is not about fundamental correctness but
rather the amount of numerical effort required to get a satisfactory result.

A related issue is that formulations of SPH with constant artificial
viscosity include a substantial shear viscosity term which leads to excess angular
momentum transport in differential rotation and subsonic turbulence
\citep{Bauer2012}.
A popular partial solution has been the \citet{Balsara1995} switch,  but it is too noisy
to strongly limit viscosity in general flows.
An effective approach is time-dependent
viscosity reduction away from shocks as first proposed by
\citet{Morris1997a}.   We note that in the past
many codes did not include this as a default option
(e.g. {\sc Gadget2} and the original {\sc Gasoline}).
\citet{Bauer2012} ran {\sc Gadget2} with
a switch like that of \cite{Morris1997a} and
achieved similar inertial ranges to grid codes (albeit with less small scale structure).
Thus we do not include excess artificial viscosity as an outstanding
problem for traditional SPH.
\cite{Cullen2010} (hereafter CD) proposed a more complex switch which
demonstrates little numerical angular momentum transport in various rotating tests.
\cite{Price2012b} and \cite{Hopkins2013} have demonstrated
excellent performance by modern SPH 
with time-dependent viscosity switches
on shearing problems including subsonic turbulence.
  
The Lagrangian nature of SPH particles means that fluid quantities do not mix at
small (sub-resolution) scales.  This would be correct for laminar
flows with limited molecular diffusion but most astrophysical flows have high
Reynolds numbers and are expected to have a turbulent cascade that
continues down below resolved scales.   For example,
\citet{Wadsley2008} showed that a lack of
turbulent diffusion led to the low entropy cores for SPH simulations of
galaxy clusters in \citet{Frenk1999}.   \citet{Shen2010} showed that
turbulent diffusion is required for realistic distributions of metals in and around
simulated galaxies.   These issues are most apparent in the presence of gravity,
where buoyancy and convection separate particles with different entropies.
Mesh codes diffuse all quantities due to
numerical diffusion errors associated with advection, which can be close to the
expected physical diffusion when the absolute fluid velocities are
small.   Problems arise in structured, high-Mach-number flows, such as in
typical disk galaxies \citep{Benincasa2013}.

A complete lack of mixing can suppress the growth of fluid instabilities, making
traditional SPH unable to capture the growth of Kelvin-Helmholtz or
Rayleigh-Taylor instabilities across discontinuous jumps in thermal energy
\citep{Price2008,Read2010}.   This can lead to dense structures failing to mix
within the correct timescales \citep{Agertz2007,Wadsley2008}.
%This can promote blobs (Kaufmann) [which may or may not be real in
%nature note!] but overall gas accretion rates similar
%(e.g. Governato -- some paper).

Every simulation of turbulent flows should include
explicit turbulent transport terms, of which the leading order term is turbulent diffusion.
 %LES in astro -- someone like wolfram schmidtdd 
\citet{McNally2012} showed that explicit diffusion can be
required to prevent numerical noise overtaking explicit perturbations such as in
the Kelvin-Helmholtz instability.  The most complex schemes explicitly evolve
unresolved turbulent energy \cite[e.g. Large-Eddy Schemes,][]{Germano1991,Schmidt2006}.  
For these reasons, in {\sc Gasoline2} we favour the idea of turbulent
diffusion based on local velocity shear to mix fluid quantities.  We
note that approaches to mixing such as artificial conduction are also
effective but can have side effects in
gravitational fields, making an approach based on local velocity gradients
more generally applicable \citep{Price2008}.

Numerical surface tension exacerbates problems with mixing, realistic
treatment of multiphase flow, and two-phase instabilities.    
Surface tension arises from the formulation of pressure forces within
traditional SPH and is related to inconsistencies in the pressure and
force profiles at constant-pressure density jumps.  Proposed solutions include
estimating pressure directly rather than density \citep{Ritchie2001}
and higher numbers of neighbours \citep{Read2010}.  The large number of
neighbours and associated numerical expense prevented widespread
adoption of these methods.  Strong artificial conduction can
reduce the effect of surface tension by removing sharp density gradients \citep{Price2008}.
\citet{Ritchie2001} promoted the idea of smoothing to calculate
pressure directly and then deriving density from pressure.
\citet{Hopkins2013} employed Pressure-Entropy 
style SPH, combining this idea with an entropy evolution equation.  He demonstrated that
substantial improvement could be achieved with modified
pressure-based formulations without large neighbour numbers.
{\sc Gasoline2} uses a new pressure force formulation to achieve similar results.

In some regimes, SPH errors scale sub-linearly with resolution due to the ``E0-error'',
which is associated with the re-gridding noise that keeps SPH particles in a glass-like
configuration in evolved flows \citep{Read2010}.  The AREPO method
\citep{Springel2010a} helps resolve this by separating the motion of volume
elements from that of the fluid at the cost of not being a strictly
Lagrangian method.   SPH can reduce this error with large
neighbour numbers.  However, traditional SPH
kernel functions are unable to use neighbour numbers larger than
$N_{\rm smooth}\sim50$ due to a pairing instability which effectively limits the
resolution to $N_{\rm smooth} < 50$.  This issue was fully
resolved by \citet{Dehnen2012}, who showed the origin of the pairing
and that \cite{Wendland1995} smoothing kernels do not suffer from it.

Some issues have only come to the fore due to recent enhancements in
resolution.  For example, highly resolved Sedov blasts (Supernova
explosions) with temperature jumps exceeding factors of 100 have only
been a feature of galaxy simulations in recent years.  This led to the
realisation that hydrodynamics codes must avoid rapid spatial variations
in individual time steps \citep{Saitoh2009}.

It should be noted that specific, idealized tests can still be problem
free for specific subsets of these modern SPH features.  Our goal has
been to develop a code where a single set of parameters and algorithms
gives good results on all tests relevant to our application areas.

Modern SPH codes include features 
that alleviate the issues present in traditional SPH.  As noted earlier,
limiting numerical viscosity is important and the \cite{Cullen2010} approach
is commonly considered the current state-of-the-art.
The use of new kernels rather than
the traditional SPH M4 spline kernel allows for more neighbours and higher accuracy
(e.g. the {\sc Gasoline}
derived OSPH, \citealt{Read2010}).
Many codes use newer forms of the SPH equations of motion
(e.g. the {\sc Gadget}-derived PSPH, \citealt{Hopkins2013}).
Modifying pressure forces and using explicit mixing terms (e.g. {\sc gcd+},
\citealt{Kawata2013}) both reduce surface tension problems.
Many codes include a combination of these features (e.g.  the {\sc
  Gadget} derived codes of \citealt{Hu2014} and \citealt{Beck2016}).

The paper presents  a complete description of all the methods now
present in {\sc Gasoline2} making it a state of the art, modern SPH
code in section~\ref{sec:method}.  We demonstrate 
the accuracy and performance of the resulting code in section~\ref{sec:tests}.
{\sc Gasoline} has been continually upgraded, including turbulent diffusion
from 2008 and geometric-average-density pressure forces (see
section~\ref{sec:forces}) in 2011.  These were sufficient to address
major issues with fluid instabilities (as shown in
section~\ref{sec:tests}).  The primary feature that was lacking until
recently was CD-like time-dependent viscosity.  {\sc Gasoline2} also has unique
features to limit viscosity in convergent flows which previously
published viscosity limiters fail to do.  We provide detailed
motivations for our preferred formulation and demonstrate the benefits
with specific tests.  We conclude by discussing future directions for
Lagrangian hydrodynamics in section~\ref{concl}.

\subsection{{\sc Gasoline}, { \sc Pkdgrav} and {\sc Changa}}

All versions of the {\sc Gasoline} code have been built upon the {\sc pkdgrav} parallel
tree code \citep{Stadel2001}.  
{\sc Gasoline2} retains the core architecture of {\sc pkdgrav}.  This includes
the KDK integration scheme, gravity solver, and tree structure.  
In 2005, it was cutting edge with
comparable or superior performance to other codes (e.g. {\sc Gadget2}  and
3, \citep{Springel2005b}).
However, new programming approaches were needed
to take advantage of systems with thousands of cores.  The core gravity and parallel
routines in {\sc Gasoline} stopped active development around 2008 in
response to the plans for new, scalable parallel alternatives.
{\sc Pkdgrav} continued its development as an N-body only code, culminating
in the {\sc Pkdgrav3} code \citep{Potter2016}.
{\sc Pkdgrav3} is optimized for solving gravity on large parallel systems with GPUs and has
run over a trillion particles.

Combining both hydrodynamics (SPH) and gravity is challenging for
parallelization with distributions of work and data highly clustered
in both space (large density contrasts) and time (multiple time steps).
These distributions also differ for different species (e.g. dark
matter, stars and gas).  An effective approach is inherently parallel
languages such as {\sc Charm++} \citep{Kale1993} based on {\sc C++}.
The {\sc Charm++} execution
model incorporates functions as asynchronous, concurrent tasks, work
migration and internal task scheduling.  These are managed without the
direct intervention of the programmer.  This model allows simulations to adapt to
work imbalance, node failures and heterogeneous architectures at run
time (see also: \citealt{Theuns2016}).  {\sc Changa}
is a TreeSPH code written in {\sc Charm++} that incorporates many of the
gravity code improvements of {\sc pkdgrav} and all of the SPH methods
implemented in {\sc Gasoline2}.  We refer the reader to \cite{Menon2015} for
details of code design and performance.  The most important
difference is the use of an oct-tree rather than a binary KD-tree for
gravity.  This changes the distribution of the gravity force errors
and thus calculations involving gravity are not identical using
{\sc Gasoline2} and {\sc Changa}.  The gravity improvements include fast
multipole methodology \citep{Potter2016},
so that {\sc Changa} is also intrinsically faster than
{\sc Gasoline2} for N-body simulations.  On the other hand, SPH is based on
neighbour lists and it is thus possible to achieve identical SPH
densities and forces to round off.  Over time, even round off-level differences 
accumulate and lead to noticeable differences at the particle level.

{\sc Gasoline2} and {\sc Changa} have been developed together with the objective
of maintaining identical SPH methods and results.  In most functions the
code has been directly copied.  We will focus on the
current SPH methods employed in these codes in
section~\ref{sec:method} and present results for standard test
problems in section~\ref{sec:tests}.  The results presented here were run with
{\sc Gasoline2}.    {\sc Changa} produces essentially identical results.

\section{Method}
\label{sec:method} 

\subsection{SPH Estimators, Kernels and Smoothing Lengths}

All SPH codes are based on local summations over neighbours.  For some
fluid quantity, $f_j$, known at particle positions, $\vec{x}_j$, we
get an SPH smoothed estimate as follows,
\begin{eqnarray}
f_{i,{\rm smoothed}} = \sum_{j} f_j {\rm W}(\vec{x}_i-\vec{x}_j,h_i,h_j).
\label{sum}
\end{eqnarray}
\noindent where ${\rm W}$ is a general kernel function and $h_j$ is a
per-particle smoothing length, indicative of the range of interaction
of particle $j$.  It is standard practice to directly refer to
quantities such as density that are derived this way with the {\it
  smoothed} qualifier.  For momentum and energy conservation a symmetric
expression is required in the force terms, which led many authors to
symmetrize the kernel in all summations.  Starting with \cite{Springel2002}
it became common to use un-symmetrized or gather estimates for
non-force terms such as the density.  For the density and other per-particle 
quantities {\sc Gasoline2} uses the gather (i.e. one-sided) estimate
which is equivalent to,
\begin{eqnarray}
{\rm W}(\vec{x}_i-\vec{x}_j,h_i,h_j) &=& \frac{1}{h_i^3} W(\frac{r_{ij}}{h_i}).
\label{kernel}
\end{eqnarray}
\noindent where $\vec{r}_{ij} = \vec{x}_i-\vec{x}_j$ and $r_{ij}$ is
its magnitude. 
For the specific kernel function, $W(q)$, we use the Wendland kernels
\cite{Wendland1995} which do not suffer from a pairing instability.  We also follow 
\citealt{Dehnen2012}) in adjusting the
kernel weight at particle $i$ to correct for the small bias
in the density estimator.  {\sc Gasoline2} can be run using
other kernels such as the traditional cubic spline \cite{Monaghan1992}.

We define the smoothing length, $h_i$, as half the distance to the
furthest neighbour, consistent with its use in the original SPH
literature \citep[e.g.][]{Monaghan1992}.  To find the smoothing length we find an
exact number of neighbours.  Our standard neighbour number (as used on the
tests below) is 200.  Historically, many {\sc Gasoline} simulations were
run with 32-64 neighbours.  {\sc Gasoline} and {\sc Changa} employ search
algorithms based on priority queues which efficiently return complete
one-sided neighbour lists and do not require iterations.

Gadget2 \citep{Springel2002} and many other codes weight neighbours in the same way
as the density sum and require a small number of iterations to find
${h_i}$.  \cite{Hopkins2013} pointed out that the weightings used for
neighbour selection constraints can be completely generalized
independent of the density expression.  A fixed neighbour count
corresponds to a uniform weighting.  It should be noted that some
codes (e.g. {\sc Gadget2}) do not allow large deviations from the typical
number of neighbours.
A fixed neighbour count constraint can be incorporated into SPH
frameworks derived from a Lagrangian \citep{Nelson1994} but such
frameworks are not used here.  Consequences and implications for the
scheme are discussed in section~\ref{sec:forces}.

Density in {\sc Gasoline2} thus only depends on quantities known at particle $i$ and
the positions of its neighbours,
\begin{eqnarray}
\rho_i &=\sum_{j} m_j W(r_{ij},h_i).
\label{density}
\end{eqnarray}
\noindent This is computationally efficient, particularly when only a
subset of particles requires updated densities.  
% Avoid reference to DI SPH -- they vary particle masses
%In the case of uniform
%particle masses this density expression is equivalent to how
%density-independent SPH selects neighbours \cite{saitohmakino?} as 
%$m_j$ may be taken outside the sum.  %check this

\subsection{SPH gradients}\label{gradients}

The one-sided gradient of the kernel is, 
\begin{eqnarray} 
\nabla_i {\rm W}(r_{ij},h_i) = (\vec{x}_i-\vec{x}_j) \frac{1}{h_i^5} W'(q),
\end{eqnarray}
\noindent where $W'(q) = \frac{1}{q} \frac{dW}{dq}$ and $q=r_{ij}/h_i$.

Most modern codes use local gradients in velocity to estimate
diffusion and limit dissipation terms associated with artificial
viscosity.  These may be calculated at the same time as density via
one-sided estimates.  For example, the divergence and curl of the
velocity field $\vec{v}_i$ at the location of the particle can be
constructed directly from components of the velocity gradient tensor,
\begin{eqnarray}
{\bf V}_{\alpha\beta} = \frac{\partial v_{\alpha}}{\partial x_{\beta}}.
\end{eqnarray}
\noindent where $\alpha$ and $\beta$ run through the Cartesian axes.

We employ a straightforward estimator for the velocity gradient tensor is as follows,
\begin{eqnarray}
\left. {\bf V}_{\alpha\beta} \right\rvert_i &=
\frac{\sum_{j} (v_{\alpha\,i}-v_{\alpha\,j})
  (x_{\beta\,i}-x_{\beta\,j}) m_j \frac{1}{h_i^5} W'(\frac{r_{ij}}{h_i})}
{\frac{1}{3} \sum_{j} r_{ij}^2 m_j \frac{1}{h_i^5}
  W'(\frac{r_{ij}}{h_i})}
\label{velocitygradient}
\end{eqnarray}

\noindent The numerator is a standard SPH gradient estimate and the
denominator expression provides the improvement.  This estimator is
exact in the case of uniform contraction or expansion
and corresponds to CD equation B1.
CD point out that it is biased if there are significant density gradients.  However,
we find that this bias is modest compared to the effects of particle noise on the estimator.
For the purposes of estimating artificial dissipation the simpler form is
sufficient.  In particular, we see no difference on test problems.

For symmetrized force terms we require a symmetric gradient of the
kernel, $\nabla_i W_{ij}$,
\begin{eqnarray}
\bar{\nabla_i W_{ij}}=\frac{1}{2}
f_i \nabla_i W(r_{ij},h_i) + \frac{1}{2} f_j \nabla_j W(r_{ij},h_j).
\label{kernelavg}
\end{eqnarray}
The term,
\begin{eqnarray}
f_i= \frac{\sum_{j} \frac{m_j}{\rho_i} r_{ij}^2 W'(\frac{r_{ij}}{h_i})}
  {\sum_{j} \frac{m_j}{\rho_j} r_{ij}^2 W'(\frac{r_{ij}}{h_i})},
\label{fcorr}
\end{eqnarray}
is a correction of order unity that can be compared to the $f_i$
term in \cite{Springel2002} and is discussed in section~\ref{entropy} below.
The symmetrized expression simply reverses sign on the interchange of $i$ and $j$.
In practice we accumulate force terms via two separate summations to
particles $i$ and $j$ with different smoothing lengths in each case.  

\subsection{Hydrodynamical Equations}\label{sec:forces}

SPH is an approach to solving Euler's equations, given in Lagrangian form as,
\begin{eqnarray}
\frac{d\,\rho}{d t} &=& -\rho \nabla \cdot \vec{v} \label{e1}
\\
\frac{d\,\vec{v}}{d t} &=& -\frac{\nabla P}{\rho} + \vec{g}\label{e2}
\\
\frac{d\,u}{d t} &=& -\frac{P}{\rho} \nabla \cdot \vec{v} +
\Gamma - \Lambda, \label{e3}
\end{eqnarray}
\noindent where $\rho$, $\vec{v}$ and $u$ are the fluid density,
velocity and internal energy respectively, and the three equations
express mass, momentum and energy conservation respectively.  
$\vec{g}$ represents body forces such as gravity.  
To close the system we need an equation of
state such as $P=(\gamma-1) \rho u$ for an ideal gas, where $\gamma$ is
the ratio of the specific heats.   The precise form
depends on the complexity of the fluid being simulated and is closely
connected to heating, $\Gamma$, and cooling, $\Lambda$, terms which
are generally complex functions of density, temperature and composition.
In what follows we omit body forces and heating and cooling terms.

Equation~\ref{e1} is automatically satisfied in SPH by equation~\ref{density}.

The momentum equation is expressed as,
\begin{eqnarray}
\frac{d\,\vec{v}_i}{dt}& = & -\sum_{j} m_j 
\left({\frac{P_i+P_j}{\rho_i \rho_j}+\Pi_{ij}}\right) \bar{\nabla_i W_{ij}}, 
\label{dvdt}
\label{sphmom}
\end{eqnarray}
\noindent where $P_j$ is pressure, $\vec{v}_i$ velocity and $\Pi_{ij}$
is an artificial viscosity term, discussed in
section~\ref{sec:av}.  This {\it Geometric Density Average Force} (GDF)
expression is unique to {\sc Gasoline}.
It is a member of the general family of pressure gradient expressions
described in \citet{Monaghan1992} (equation 3.5, setting $\sigma=1$).
Choosing this form was inspired by ideas
presented in \cite{Ritchie2001} 
aimed at minimizing errors in strong density gradients.  It has the
desirable property that it closely mimics the form of
equation~\ref{e2}, where the local density ($1/\rho_i$) can be taken out front and the two
pressures have the same multipliers.  It is found to minimize surface
tension type effects highlighted by \cite{Agertz2007} as is
demonstrated in the tests in section~\ref{sec:tests}.  It also
naturally complements the form of the energy equation.

The internal energy equation uses a form similar to that
advocated by \cite{Evrard1988} and \cite{Benz1990},
\begin{eqnarray}
\frac{d\,u_i}{dt}& = & \sum_{j} m_j \left( \frac{P_i}{\rho_i \rho_j} + \frac{1}{2} 
\Pi_{ij} \right ) \vec{v}_{ij} \cdot \bar{\nabla_i W_{ij}} , 
\label{dudt}
\end{eqnarray}
\noindent where $u_i$ is the internal energy of particle $i$ and
$\vec{v}_{ij}=\vec{v}_i-\vec{v}_j$ is the velocity vector difference
between the particles $i$ and $j$.  Note the re-use of the Geometric
Density Average.
In concert with equation~\ref{dvdt} this form conserves energy exactly in each
pairwise exchange.  For a purely adiabatic gas this form is equivalent
to $P/\rho$ multiplied by an SPH estimate for the divergence and thus
closely follows the physical equation~\ref{e3}.

\subsubsection{Entropy conservation}\label{entropy}

SPH models the fluid system via pairwise exchanges of momentum and
energy between particles.  These exchanges must be symmetric to
achieve momentum, angular momentum and energy conservation.  This
conservation is
only exact in the limit as the time step approaches zero and is
otherwise dependent on the quality of the integrator.  Exact momentum
conservation can be achieved with any integrator and fixed time steps.
If the SPH force expressions are derived from a Lagrangian (as in
\cite{Springel2005a} or \cite{Hopkins2013}), symmetry is automatic and the
form of the equations is fixed.  This approach limits ones ability to
adjust the form to minimize problems such as surface tension-like
behaviour at sharp jumps in the pressure.  On the other hand, the
Lagrangian approach provides accurate entropy conservation
\cite{Springel2002}.  It is instructive to show how this comes about.
In the absence of shocks, sources or sinks, the
entropy function, $A(S)=u/\rho^{\gamma-1}$ is constant on a mass element.  For this to
remain true, density and internal energy must evolve consistently so
that,
\begin{eqnarray}
-\nabla \cdot \vec{v} =
\frac{1}{(\gamma-1) u} \frac{du}{dt} = 
\frac{1}{\rho} \frac{d \rho}{dt}.
\label{entropydiv}
\end{eqnarray}
For an arbitrary SPH formulation, even if energy is manifestly
conserved, there may be a systematic offset in the estimate for
$du/dt$ so that when density changes by orders of magnitude the
thermal energy does not change appropriately and entropy is not
conserved.  One approach is to evolve entropy directly but this can
create energy errors.  The issue can be addressed directly by
correcting the momentum and energy equations to exactly respect
equation~\ref{entropydiv}.  In formulations derived from a
Lagrangian, this correction, called $f_i$ in \cite{Springel2002}, is
automatically introduced by the Lagrange multipliers.

Our internal energy equation (eqn.~\ref{dudt}) is only dependent on the local particle
pressure making it closer to the physical energy equation~\ref{e3} than most
alternative forms.  As a consequence entropy is closely conserved.

However, it possible to explicitly estimate a correction for it (or 
any SPH formulation) by comparing the estimate for $\nabla\cdot
\vec{v}$ on the left-hand side of eqn~\ref{entropydiv}, arising from
the internal energy (eqn.~\ref{dudt}), to that on the right side in
eqn.~\ref{entropydiv} derived from eqn.~\ref{density},
\begin{eqnarray}
\frac{1}{\rho_i} \frac{d\,\rho_i}{dt}& = & \sum_{j} \frac{m_j}{\rho_i} \frac{W'}{h_i^5} 
\vec{v}_{ij} \cdot \vec{r}_{ij} \\
& - & \sum_{j} \frac{m_j}{\rho_i} r_{ij}^2 \left[ \frac{\,W'}{h_i^6} + \frac{3\,
    W}{h_i^4}\ \right] \frac{dh_i}{dt}.
\label{drhodt}
\end{eqnarray}

\noindent The term the square brackets in equation~\ref{drhodt} comes from
$\partial \rho/\partial h$ and was used to derive correction factors
$f$ in \cite{Springel2002}.  Such correction terms are commonly referred to
as ``grad h'' terms.  Under the idealized integral approximation of the
sum, this term is identically zero.  We find no systematic effect on
entropy integration arising from it for the formulation presented
here.  In cases where severe errors occur (large amounts of expansion
or contraction),
% note in this case $\rho \propto h^{-3}$
the dominant effect is from the first term above.  In eqn.~\ref{dudt},
$\rho_j$ replaces $\rho_i$ in this term.  This gives the form of
$f_i$ presented in eqn.~\ref{fcorr}.

Using the correction in equation~\ref{fcorr} ensures
that the ratio of the numerical estimates of the two key terms in
equation~\ref{entropydiv} does not systematically vary from unity.
Therefore entropy is closely conserved.  This is demonstrated in the
test section below, such as for the Evrard test where Gasoline exactly
follows the correct entropy evolution prior to the shock and in the
Sedov test.   It can also be easily demonstrated by simulating an
expanding sphere for many expansion factors and showing that the
entropy is conserved.  Traditional SPH forms without correction terms
(e.g. the original {\sc Gasoline}, \citealt{Monaghan1992}) see systematic
variations in the entropy.

\subsection{Shocks and Artificial Viscosity}\label{sec:av}

All hydrodynamics codes require extra numerical dissipation to broaden
dissipative shocks whose natural width is smaller than their
resolution element.  Without this they are numerically unstable.  This
regime is common in astrophysics where molecular viscosity is very
low.  Codes based on piecewise reconstructions and Riemann solvers
(e.g. \citealt{Woodward1984,Springel2010a,Hopkins2015}) do this through
slope-limiters, making the solver first order in shocks and thus
locally equivalent to the original \cite{Godunov1959} scheme.  Central
schemes \citep[e.g.][]{Kurganov2000} apply slope limiters but add
dissipation in a manner similar to artificial viscosity.  Direct
artificial viscosity is the dominant approach in SPH.   In all
schemes, the goal is to
minimize dissipation away from shocks.  The original SPH form
\citep{Monaghan1992} is as follows,
\begin{eqnarray}
\Pi_{ij} = \left\{{ \begin{array}{ll}
\frac{-\alpha\frac{1}{2}(c_i+c_j)\mu_{ij}+\beta\mu_{ij}^2}{\frac{1}{2} (\rho_{i}+\rho{j})}
& {\rm for~}\vec{v}_{ij}\cdot\vec{r}_{ij}< 0,\\
~0 & {\rm otherwise}, \end{array}}\right.\\
{\rm\ where\ }
\mu_{ij} = \frac{h(\vec{v}_{ij}\cdot\vec{r}_{ij})}{\vec{r}_{ij}^{\,2}+0.01 (h_i+h_j)^2},
\label{artifvisc}
\end{eqnarray}
\noindent where $c_j$ is the sound speed of particle $j$. 
$\alpha=1$ and $\beta=2$ are standard values for parameters
representing shear and Von Neumann-Richtmyer (high Mach number)
viscosities respectively.  We have implemented this and also the signal speed
variant of the artificial viscosity \citep{Morris1997b}
 in {\sc Gasoline2},
\begin{eqnarray}
\mu_{ij} = \vec{v}_{ij}\cdot\hat{r}_{ij},
\end{eqnarray}
The viscosity is then effectively a function of the signal speed $v_{\rm sig, ij} =
\frac{c_i+c_j}{2} - \min(\vec{v}_{ij}\cdot \hat{r}_{ij}, 0)$ which better approximates a local
wave-speed of the Euler equations.
% We use beta = 3/4 normal beta in this case -- no reference in the
% code as to why

%NOTE: Cullen and Dehnen different definition of mu 2x bigger due to H
%vs h

In past work with {\sc Gasoline} we used the multiplicative \cite{Balsara1995}
limiter to reduce artificial viscosity,
\begin{eqnarray}
\xi_{\rm Balsara} &=& \frac{\mid\nabla\cdot\vec{v}\mid}
{\mid\nabla\cdot\vec{v}\mid+\mid\nabla\times\vec{v}\mid}.
\end{eqnarray}
\noindent This limiter is effective in non-shocking, shearing
environments but is otherwise very noisy and is thus half on ($\sim 0.5$) in most
of the simulation volume.

\cite{Morris1997a} suggested a time-dependent limiter implemented via 
evolving the viscous parameter $\alpha$, such that $d\alpha/dt =
\max(0,- \nabla \cdot \vec{v}) - \alpha/\tau$.  This increases in
compressions and otherwise slowly decays with $\tau \sim 0.1\,h/c$.  The
decay acts somewhat like a smoothing in time and helps compensate for
noise.  CD pointed out that $\alpha$ increases
gradually, potentially too slowly for strong shocks.  A second problem
is that it cannot differentiate between compressions and shocks.
CD suggested instantly setting $\alpha$ to a
maximal value based on the time derivative of the divergence,
$-d(\nabla \cdot \vec{v})/dt$.  This approach allows for rapid activation
of artificial viscosity.  It is noisy away from shocks,
particularly when unavoidable particle reordering is occurring such as
during density changes.  This tends to make the limiter also partly on
in much of the volume (e.g. $\alpha \sim 0.15$ between the rarefaction and
the contact discontinuity in shock tubes in CD
fig.~11).  Thus the fact that CD chose not to apply a
minimum $\alpha$ may be a moot point.
As noted by \cite{Rosswog2015} artificial
viscosity in disordered regions could be of net benefit and provide a
better overall solution if it helps limit velocity noise.  For this
reason a noise-sensitive estimator can still perform well on many tests and
Rosswog advocated for an explicit noise-based limiter.

A key problem with any viscosity approach based on divergence
(including the CD approach) is that it
creates viscosity in uniform compression (as seen in the left panel of their
fig.~15 where $\alpha \sim 0.5$ everywhere during the collapse test of
section~\ref{test:evrard}).  
The usual indicator for shocks is $\nabla \cdot \vec{v}$.  For a
spherical flow, $\nabla \cdot \vec{v} = \frac{1}{r^2}\frac{\partial
  r^2 v_r}{\partial r} = \frac{\partial v_r}{\partial r} - 2
\frac{v_r}{r}$.  In spherical collapse,
divergence is commonly dominated by the second term
and is not a reliable shock indicator.  One
could look for the most negative eigenvalue but in the collapse test,
negative, non-radial eigenvalues dominate away from the shock.

A reliable indicator is the velocity gradient in the direction of the
pressure gradient.  This can be used in any geometry.  We estimate the
pressure gradient direction and velocity gradient using the local,
one-sided estimator of section~\ref{gradients} at the same time
density is calculated, 
\begin{eqnarray}
    \nabla P &=& (\gamma-1)\sum_j m_j u_j \nabla_i {\rm W}(r_{ij},h_i), \\
\hat{n} &=& \frac{\nabla P}{|\nabla P|}, \\
\frac{d\,v}{dn} &=& \sum_{\alpha,\,\beta} n_{\alpha} V_{\alpha\beta} n_{\beta}.
\end{eqnarray}
\noindent The result, $\frac{d\,v}{dn}$, is an accurate local scalar
shock indicator in the spherical collapse case but is still triggered
in uniform compression.  For this reason we subtract off one third the
divergence (if it is negative) to ensure that the $\hat{n}$ direction
is playing a dominant role in the local compression.  The resulting
{\it gradient based} shock detector,
\begin{eqnarray}
D &=& \frac{3}{2} \left (\frac{d\,v}{dn} +
  \max(-\frac{1}{3} \nabla \cdot \vec{v}, 0 ) \right).\label{dvds}
\end{eqnarray}
\noindent $D$ takes on the extremal value of $\frac{dv}{dn} = \nabla \cdot
\vec{v}$ in the case of a shock but is only negative when the
compression normal to the shock is more than one third the overall
compression.  Other similar forms could be used.  This indicator is
also effective in the context of a Balsara-type limiter, replacing the
divergence.

The general idea of limiters is to have a comparator that measures
non-shocking flow gradients to combine with the shock indicator.
\cite{Dehnen2016} noted that limiters such as that of
\cite{Balsara1995} that use the curl as a comparator lead to different shock
properties in rotating systems.  This motivated
CD to use the magnitude of the trace-free shear
tensor, ${\bf S}_{\alpha\beta}= 1/2 ({\bf V}_{\alpha\beta}+{\bf
  V}_{\beta\alpha}) - \delta_{\alpha\beta}\,\frac{1}{3} \nabla \cdot \vec{v}$  as the
comparator.  This is zero for pure rotation but still detects shear
(e.g. differential rotation).  Removing the trace makes it zero in
isotropic compression or expansion which limits its ability to prevent
viscosity there.  In addition, in a strong shock the tensor is
dominated by a single eigenvalue.  If we align the shock normal with
the x-axis then only the ${\bf S}_{xx}$ term is large and negative.
Subtracting the trace makes the other two diagonal elements non-zero.
Then the norm $|{\bf S}| = \sum_{\alpha,\,\beta} {\bf
  S}_{\alpha\,\beta}^2$ is non-zero which undesirable.  If we keep
the trace and use ${\bf T}_{\alpha\beta}= 1/2 ({\bf V}_{\alpha\beta}+{\bf
  V}_{\beta\alpha})$, then an excellent overall indicator is
$\frac{D}{|{\bf T}|}$

Our overall scheme to limit artificial viscosity is modeled on that of
CD and is as follows,
\begin{eqnarray}
\alpha_{{\rm loc},\,i} &=& \alpha_{\rm max} \frac{A_i}{A_i + v_{sig,i}^2}, {\rm \
  where\ } \\
A_i &=& 2\, h_i^2\,\xi_i \max (-\frac{dD}{dt}, 0).
\end{eqnarray}
The only change here is in the definition of $A_i$ which is constructed with
the one-dimensional velocity derivative of equation~\ref{dvds}.   This combined
with the different $h$ definition in CD led us to boost the $A_i$ term by 2.  For the tests
shown in this work we employ $\alpha_{\rm max} = 2$.
Experiments with strong shocks indicate higher
peak $\alpha$ values may be needed.  Our parameter values were chosen based on
the test results shown in section~\ref{sec:tests}.  We also note that we
use the gradient estimator of equation~\ref{velocitygradient}
rather than the full treatment in the appendix of CD. 

Whenever $\alpha$ is less than
$\alpha_{\rm loc}$, it is set directly to that value, otherwise,
\begin{eqnarray}
\frac{d\,\alpha_i}{dt} &=& (\alpha_{{\rm loc},\,i}-\alpha_i)/\tau_i, 
{\rm \ where\ } \tau_i = \frac{h_i}{0.2 c_i}.
\end{eqnarray}
\noindent This decay is a little faster than that of CD.  We have
experimented with minimum $\alpha$ values but find that noise often
contributes sufficient base $\alpha$ value where required.  A more in
depth examination is presented in section~\ref{tests:gresho} in the
discussion of the Gresho-Chan test.

For the limiter we use,
\begin{eqnarray}
\xi_i &=& \left(\frac{1-R_i}{2}\right)^4  {\rm \ where\ } \\
R_i &=& \frac{\sum_j m_j D_j/|{\bf T}|_j W_{R,\,ij}}{\sum_j m_j W_{R,\,ij}}.
\end{eqnarray}
\noindent $R_i$ is constrained to be in the range $[-1,1]$.
Formally $-1 \leq \frac{dv}{dn}/|{\bf T}| \leq 1$ by construction.  However,
$D$ is a modification to $\frac{dv}{dn}$ and noise occasionally pushes
the value outside this range.  The weighting $W_{R,\,ij}$ could use the regular
kernel except that this only strongly weights a small number of central
particles and makes the estimate noisy.  We use $W_{R,\,ij} =
(1-(r_{ij}/(2\, h_i))^4$.  $\xi_i$ is zero in expansions, $\sim 1/16$ in
intermediate and noisy regions (including uniform compression) and is
maximized at 1 in shocks.

\subsection{Diffusion}

Diffusion for any scalar quantity in SPH, $A_i$, can be estimated using \citep
{Monaghan1992},
\begin{eqnarray}
\frac{dA_i}{dt}|_{\rm Diff} = -\sum_j m_j
\frac{(d_i+d_j)(A_i-A_j)(\mathbf{r}_{ij}\cdot\nabla_i
{\bar W_{ij}})}{\frac{1}{2}(\rho_i+\rho_j)\,\mathbf{r}_{ij}^2},
\end{eqnarray}
\noindent where $d_i$ is the diffusion coefficient of particle $i$.
If diffusion rates vary rapidly in space, a harmonic mean may be useful
to replace $(d_i+d_j)/2$ above, but this is not our default.

As noted in \cite{Wadsley2008}, traditional SPH lacks a mechanism to model
the diffusion that is intrinsic to unsteady flows.  Mesh codes diffuse
numerically at rates proportional to the absolute fluid velocity which make
them non-Galilean invariant \citep{Wadsley2008,Hopkins2013}.

Turbulent diffusion dominates in high Reynolds number flows where
shear can drive local fluid instabilities.  Sub-grid turbulence may be
modelled in SPH using the trace-free local shear tensor, ${\bf S}$
\citep{Wadsley2008,Shen2010}.  We estimate the turbulent diffusion
coefficient using $d_i=C |{\bf S}| h_i^2$ with a coefficient $ C \sim
0.03-0.1$. We used $C = 0.03$ here.  This estimate avoids diffusion in pure rotation and
non-shearing flows.  We use the shear tensor estimate described previously.

Every fluid scalar should diffuse, including thermal energy, $u_i$,
metals and so forth, and this is done in {\sc Gasoline2}.  Diffusion
should be considered an essential component of any SPH code unless the
target is solely laminar flows.  Incorrect results for entropy
transport (e.g. galaxy cluster entropy profiles) and hydrodynamic
instabilities are a consequence of its absence.  We demonstrate this
explicitly through the destruction of a gas blob via fluid
instabilities in section~\ref{tests:blob}.

There are other astrophysically relevant sources of diffusion such as
Spitzer conduction in hot plasmas (used in \citealt{Keller2014}).
\cite{Monaghan1987} has shown that artificial diffusion
proportional to the viscosity can be necessary in extremely strong shocks.
We not use it for the tests shown here.  We apply all sources
of diffusion within the same framework by adding to the diffusion rate
coefficient, $d_i$.

\subsection{Time integration}

{\sc Gasoline} has used Kick-Drift-Kick 2$^{nd}$ order leapfrog
integration with hierarchical powers of two time stepping since the
original implementation \citep{Stadel2001,Wadsley2004,Springel2005b}.  This approach is
symplectic only if time steps never change.  When particles change
time steps (by factors of two each time), secular drift in otherwise
conserved quantities can be introduced though this is not systematic
and the code conserves energy well overall as shown in the original
paper and confirmed in the tests.  Recent work \citep{Springel2010a,Hopkins2015}
has suggested summing the total fluxes of every pairwise momentum and energy exchange 
to allow for exact conservation of these quantities.  This
is appealing in principle but not symplectic.   Issues of this nature
can be minimized by ensuring neighbouring particles have similar
time steps.  Thus in {\sc Gasoline2}, time step criteria are applied in a
pairwise fashion.  For example, the Courant condition is applied as, 
\begin{eqnarray}
\Delta t|_{\rm Courant, i} = 0.4 \min_j \frac{\bar{h}_{ij}}{1.25\bar{c}_{ij}+0.75\left[\alpha_{ij}\bar{c}_{ij} + \beta\mu_{ij}\right]},
\end{eqnarray}
\noindent which is symmetric between the particles.  The term in
square brackets arises where artificial viscosity is on.  This form is
standard \citep[e.g.][]{Monaghan1992}.  We also employ the acceleration criterion
$\Delta t|_{\rm Acc} = 0.2 \sqrt{\frac{h}{a}}$.  

\cite{Saitoh2009} showed that large differences in time steps
between nearby particle can cause disastrous energy non-conservation.
Following their suggestion we never let a particle time step exceed 4
times the current time step estimate of any neighbour.  In
addition, if high speed events insert new neighbours with
significantly shorter time steps, we wake up the long time step
neighbours and cut short their current time step to adjust it downwards as much
as necessary.
This makes the scheme temporarily 1$^{st}$ order but closely limits
the overall error. It is critical for strong shocks such as the Sedov-Taylor
blast, which has negligible initial temperatures (see section~\ref{tests:sedov}).

We integrate heating, cooling and chemical networks using sub-cycling
(arbitrarily many sub-steps, smaller than the hydro time step).  These
equations are commonly stiff with very stringent stability
requirements. 
Most work with {\sc Gasoline} to date used solvers based on a
modified semi-implicit Bulirsch-Stoer algorithm \citep{Press1992} or the
{\sc CHEMEQ2} scheme \citep{Mott2001}.
Algorithms based on code from \cite{Press1992} can not legally be
included in a public release.
{\sc Gasoline2} also offers the option to cool using the public
Grackle package \citep{Bryan2014,Kim2014}.
These integrators uses midpoint estimates of the hydrodynamic
contributions so that the scheme is second order overall.  We apply a
time step limiter to the hydrodynamics based on overall changes to the
internal energy, $\Delta t|_{\rm u} = 0.25\ u/(\frac{du}{dt})$.  The
form of the energy equation~\ref{dudt} ensures that energies can never
go negative due to errors in PdV estimates.

Grid codes, in particular, commonly use operator splitting for
non-hydrodynamical terms (due to the modular nature of Riemann
solver approaches) and are thus first order in the energy integration when
cooling is included because the cooling timescale is typically much shorter.
Such choices can have important implications for heating
and cooling in astrophysical contexts and are relatively unexplored.

\begin{figure}
	\includegraphics[width=\columnwidth]{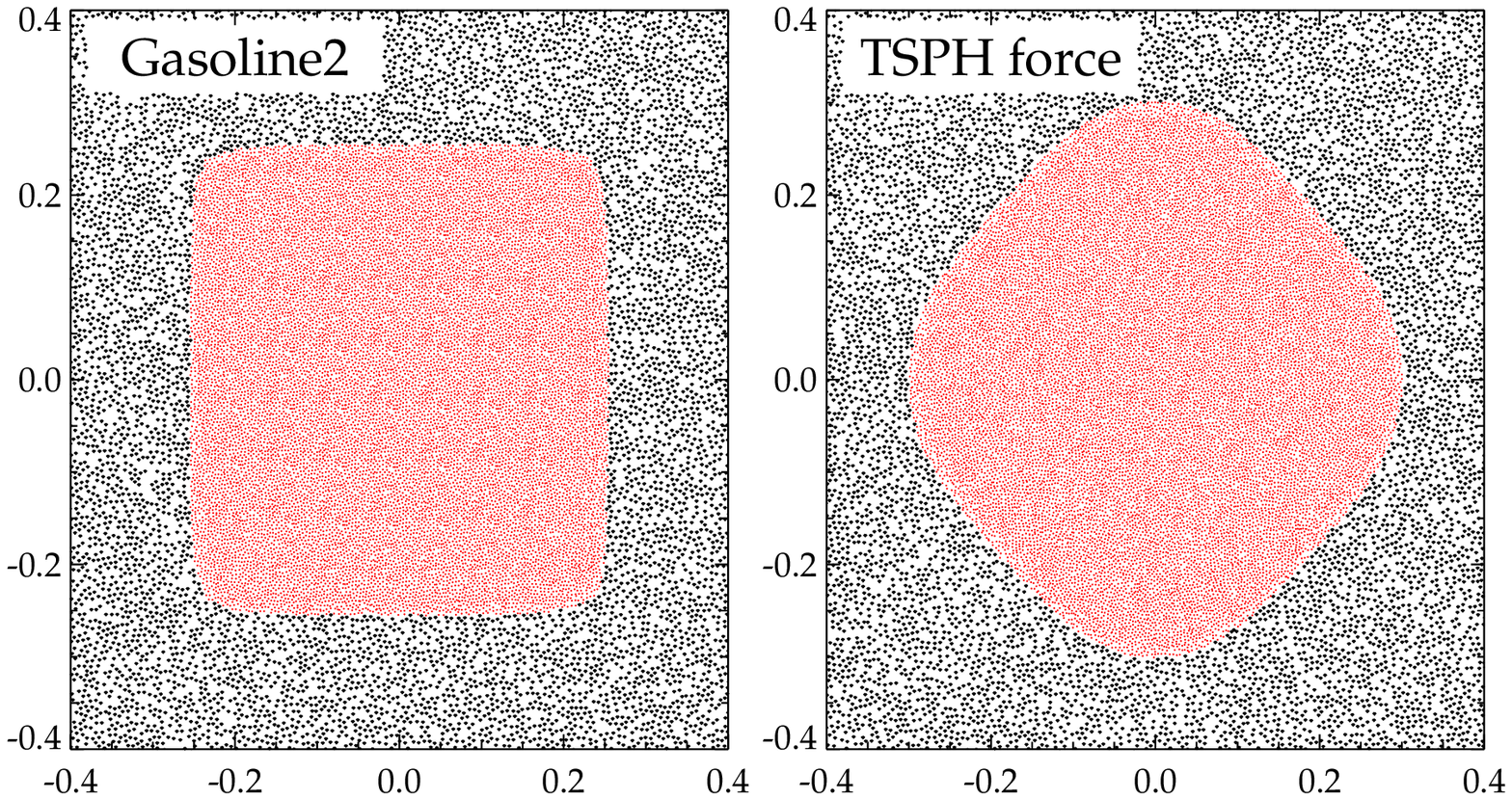}
    \caption{A comparison between {\sc Gasoline2} (which uses
      Geometric Density Forces), left,
      and a run with tradition SPH forces, right, at time t=3 on the
      square test.  The red particles are those initially within the
      square region.}
    \label{fig:square}
\end{figure}

\section{Test Problems}\label{sec:tests}

In this section we present a range of test problems on which we have
run {\sc Gasoline2} and {\sc Changa}.  All results shown were made with {\sc
  Gasoline2}.  These tests have been selected because they are
standard, they test the key problems exhibited by traditional SPH
and show how the new methods resolve them.  For our testing
we have used public initial conditions where available.
However, we also insist on using three dimensions, glass-type initial
conditions and one set of standard parameters with all modern SPH features included.
This includes optimizations such as multiple time steps which can negatively
impact conservation of energy.

Unlike grid codes, tests in one or two dimensions do not accurately
reflect how SPH will perform in three-dimensional simulations.  One
dimensional tests avoid interpenetration issues and re-gridding noise.  Two
dimensional simulations can be set up in a stable close-packed
hexagonal initial grid which minimizes noise.  A relaxed glass
configuration is the natural state for structure that is spontaneously
formed in simulations, such as through collapse or instabilities.
Glasses always have some noise and cannot be completely static.

For any numerical method it is possible to massage the results by
changing code parameters for different tests.  A well-known example of
this is slope-limiters which can be pushed into an anti-diffusive
extreme to avoid spreading at contact discontinuities
\citep{Woodward1984}.  However this choice causes numerical
instabilities in production simulations where more conservative choice
are preferred.  Thus in the following tests, we sometimes show how
different components of the method affect the results but we always
conclude with a demonstration of how the full {\sc Gasoline2} method
performs with standard parameters.  We have also avoided excessive resolution so the tests
measure practical performance.  We show the results warts-and-all,
so to speak.

\begin{figure}
    \includegraphics[width=\columnwidth]{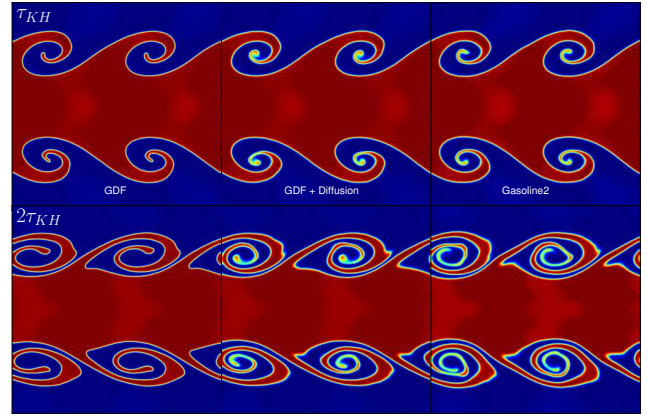}
    \caption{Kelvin Helmholtz instability test.   The
      most dramatic improvement comes from the use of Geometric
      Density Average Forces.  The left panel shows results with just
      this change from traditional SPH.  The central panel
      includes diffusion and the right panel is the full {\sc Gasoline
        2} code result.}
    \label{fig:kh}
\end{figure}

\subsection{Square Test}\label{tests:square}

This test is used to demonstrate the presence of surface tension-like
effects \citep{Saitoh2013,Hopkins2013,Hu2014,Hopkins2015}.  We perform this test in 3
dimensions starting with glass initial conditions similar to the
set-up of \cite{Hu2014}: 256$^3$ equal-mass particles in a unit cell
with a square region of 4 times higher density in $-0.25 < x < 0.25$
and $-0.25 < y < 0.25$.   Two dimensional versions and regular
lattices favour un-representative results for this test particularly.
Figure~\ref{fig:square} shows a slice through the
result at time $t=3$, with a thickness equal to the particle spacing for {\sc Gasoline2} in
its standard configuration and a version re-run with traditional force
terms.  The traditional SPH version clearly suffers from a surface
tension-like problem.  The key method improvement that gives this
result in the use of the Geometric Density Average Forces
(sec.~\ref{sec:forces}).
These results are quite comparable to those of
\cite{Hu2014} and demonstrate that pressure-entropy formulations
\citep{Hopkins2013} are not required to alleviate surface tension.  Note
that the initial condition is not in perfect force balance at the
corners which results in a small correction (rounding) there.
\cite{Saitoh2013} demonstrated that resolving this issue
completely requires variable particle masses and uniform volume
elements per particle, essentially giving up density-based adaptivity.

\subsection{Kelvin Helmholtz Instability}\label{tests:kh}

The Kelvin-Helmholtz instability is a direct test of how surface tension and
insufficient mixing suppress mixing via fluid instabilities.  We use the initial
condition based on that used in \citet{Read2010}\footnote{That IC, and those
for the blob test, are available for download at
http://astrosim.net/code/doku.php?id=home:codetest:hydrotest}.  We note that those
initial conditions feature particles aligned on a regular lattice, rather than a
glass, so we generate a set of initial conditions with identical resolution and
properties using three glass slabs.  Aside from the glassy initial particle
positions, our initial conditions are identical to those used by
\citet{Read2010}.  The domain of this IC is a rectangular slab $(L_x,L_y,L_z) =
(1,1,1/32)$.  This slab is composed of three regions, a central slab where $|y|
< 0.25$, and two slabs above and below this.  The volume is initialized with
uniform pressure, where $\rho_{\rm central}/\rho_{\rm outer} = T_{\rm outer}/T_{\rm central} =
2$.  These flows are shearing relative to each other, with each moving in
opposite directions with $v_x = 0.11\ c_s$.  A sinusoidal perturbation in $v_y$
is imposed on each of the two boundaries, defined by equation~\ref{kh_pert}
where $\delta v_y = 4v_x$ and $\lambda=0.5$.  
\begin{equation}
    \begin{split}
        v_y = \delta v_y
        [\sin{(2\pi(x+\lambda/2)/\lambda))}\exp{(-(10(y-64))^2)} - \\
        \sin{(2\pi x/\lambda)}\exp{(-(10(y+64))^2)}]
    \end{split}
    \label{kh_pert}
\end{equation}
We generate our ICs using a set of 3 glasses, rather than a grid or
lattice, to avoid artificially suppressing noise introduced by the
natural SPH re-gridding.

As shown in figure~\ref{fig:kh}, using Geometric Density Average forces is
sufficient to eliminate surface tension and allow the instability to
grow.  The addition of diffusion produces growth rates in better
agreement with high resolution simulations.   The full code version
shown in the right panels includes the variable viscosity limiter.   As can be
seen in the figure, these are are not essential for this test.   We note that alternate,
stronger forms of diffusion such as the artificial conduction of \citet{Price2008}
also improve results on this test even with more traditional pressure gradients.
However, we would argue that modified gradients with relatively low
turbulent diffusion is more generally useful.

\begin{figure*}
	\includegraphics[width=\textwidth]{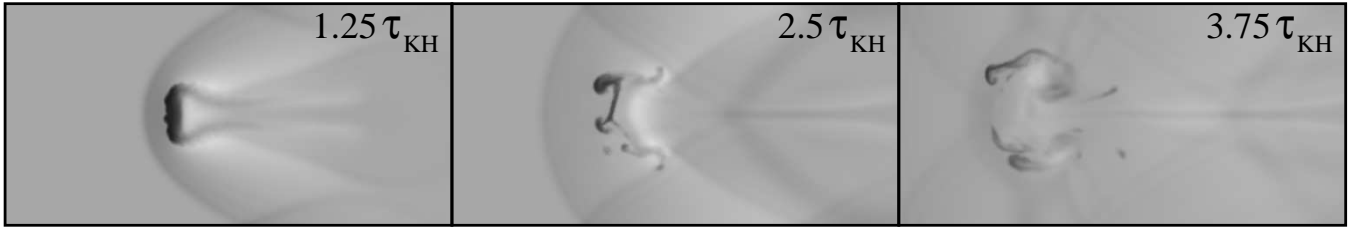}
    \caption{Agertz (2007) Blob Test: Density evolution for the full
      {\sc Gasoline2} code (log greyscale, factor of 200 in density).}
    \label{fig:blob}
\end{figure*}

\begin{figure}
	\includegraphics[width=\columnwidth]{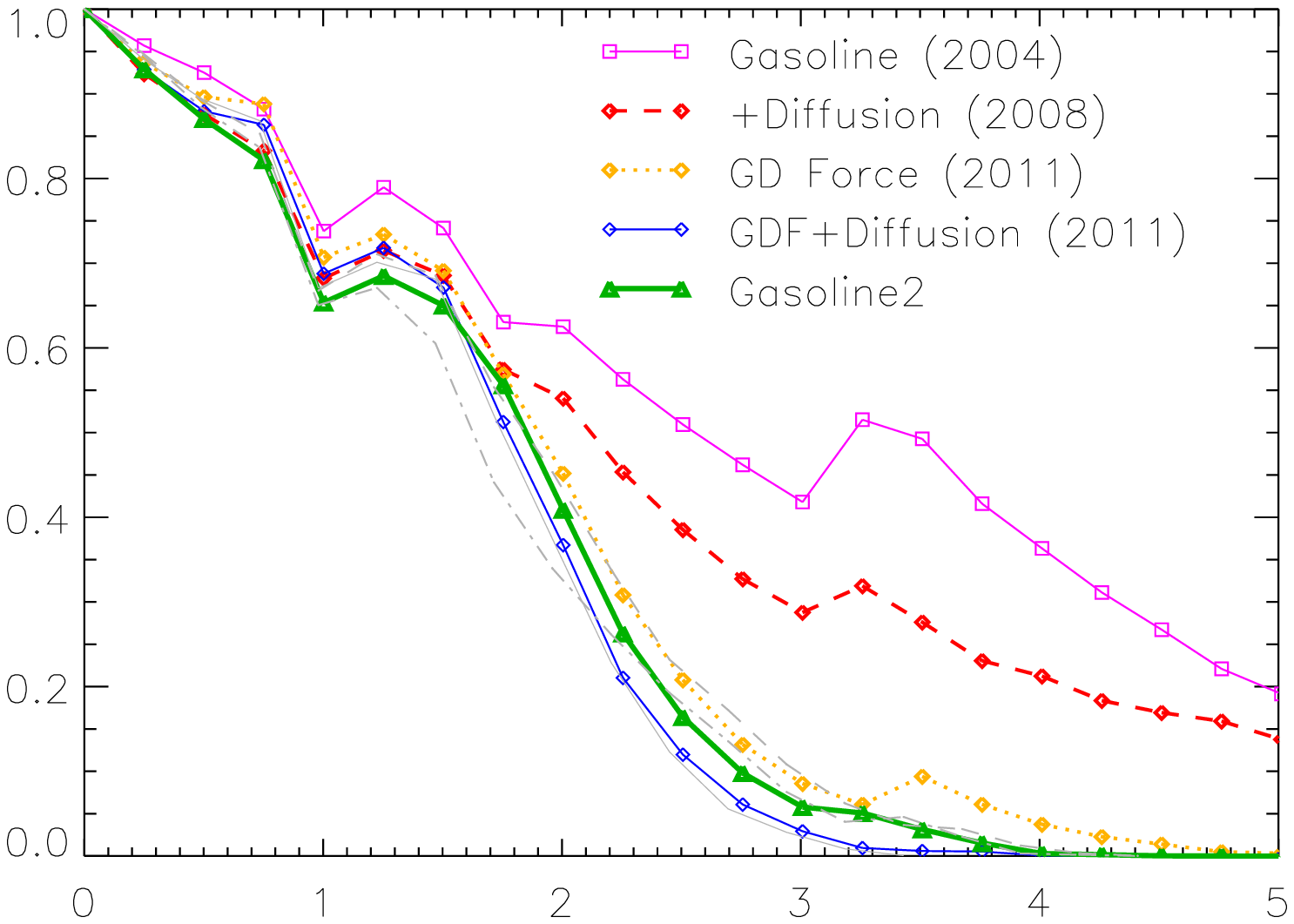}
	\includegraphics[width=\columnwidth]{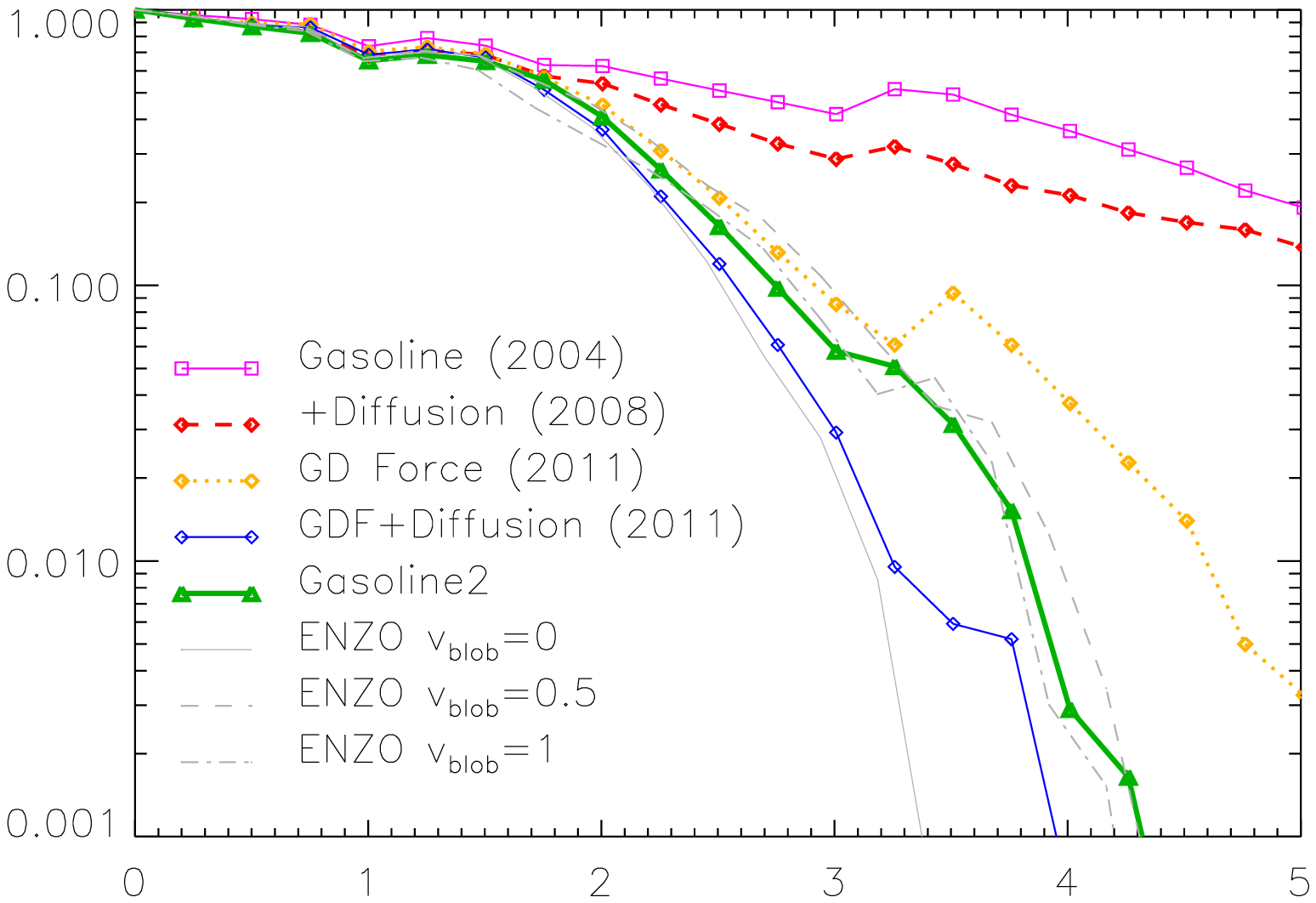}
    \caption{Agertz (2007) Blob Test:  Mass remaining versus time with
      comparison of methods.
      The top panel uses a
      linear vertical axis (as in presented in the original paper) 
      and the lower is the same data on a log axis.
      The log y-axis emphasizes how traditional SPH keeps dense
      material for long periods while mesh codes and {\sc Gasoline2} break
      the material up at an accelerating rate.  }
    \label{fig:blobmass}
% Visual of blob remaining vs. time for different runs
%% Change: put linear first, put some labels on it, others below
\end{figure}

\begin{figure}
	\includegraphics[width=\columnwidth]{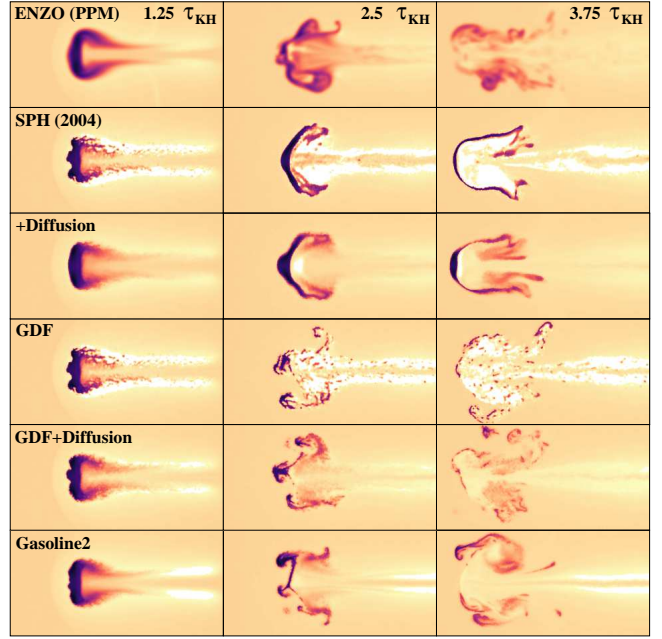}
    \caption{Blob Test results showing entropy for different methods ($T^{3/2}/\rho$ on log color scale, factor of 1000 range, bright is high).
      The top row shows standard {\sc ENZO} (AMR-PPM) results.  The second
      row is {\sc Gasoline} (2004) and the third adds just diffusion.   The
      fourth row show just the additions of the Geometric Density
      Average Force (GDF) and the fourth shows that with diffusion.
      The bottom row shows results with the full {\sc Gasoline2}
      method.  These result correspond to the curves shown in figure~\ref{fig:blobmass}.}
    \label{fig:blobentropy}
% Visual of blob in entropy -- comparing enzo, tsph, rtsph, rtsph+diff
% Place holder from old grant proposal
\end{figure}

\subsection{Blob Test}\label{tests:blob}
One of the primary difficulties traditional SPH faces is mixing of multiphase
fluids.  The ``Blob'' test \citep{Agertz2007} is designed to study 
this, by embedding a stationary dense, spherical cloud in a uniform supersonic flow.  The
cloud itself is in pressure equilibrium with the flow, and should be broken up
by the development Kelvin Helmholtz and Rayleigh Taylor instabilities as it is
accelerated up to the flow velocity.

The initial conditions for the blob test come from the Wengen 3 test
suite.
The domain is a
periodic rectangular prism, with dimensions $(L_x,L_y,L_z) = (10,10,40)$ in units of
the cloud radius, which is centered at $(x,y,z)=(0,0,-15)$. This cloud is
placed in pressure equilibrium, with $\rho_{cloud}/\rho_{wind} = 10$.  The wind
velocity is thus $v_{wind} = \sqrt{10}R_{cloud}/\tau_{KH}$.

The evolution of a density slice with the full {\sc Gasoline2} code is
shown in figure~\ref{fig:blob}.  The blob develops fluid
instabilities, breaks up and diffuses into the flow as
expected.  \cite{Agertz2007} quantified this result by examining the
fraction of the original blob mass still above 64\% of its original
density as shown in figure~\ref{fig:blobmass}.  The upper panel of
this figure shows the progression with a linear vertical axis (as in
the original paper) for different versions of the code.  
The lower panel shows the progress on log axes.
This plot shows how traditional SPH suffers with surface tension which
inhibits the initial break up.  Adding grid-scale diffusion modestly
improves the result.  

On linear axes it looks like the Geometric Density Average
Force is sufficient to achieve good results.  It follows the initial
break up well.  This period until time 3 or so is characterized
by an accelerating break up as the pieces of the cloud break-up faster
because their individual KH times are shorter.  Thus on a log plot,
the characteristic time to halve mass decreases and the slope becomes
more negative.   However, once the final pieces are close to the
resolution scale (near time 3), a new issue arises that must be
resolved through diffusion.    The full {\sc Gasoline2} and older {\sc Gasoline}
results with the new force and diffusion are slightly different and
inhabit a region of solution space similar to the range spanned by the
{\sc ENZO} results.  We have verified that other strong mixing models, such as
the artificial conduction of entropy used in {\sc PHANTOM} \citep{Price2017}
produce similar results.

The {\sc ENZO} results include three curves because fixed-grid codes are not
Galilean invariant.  The $v_{blob}=0$ curve is the standard run.   In
$v_{blob}=1$ the blob moves to the left and the background flow is
static and for $v_{blob}=0.5$ each had half the motion.  SPH gives
identical results in these three cases.   However,  for grid codes
there is faster breakup when the flow moves rapidly relative to the
grid.  This result emphasizes that there is no unique answer to this
problem.  It is sensitive to initial conditions, method choices and
effective resolution.  Though the initial condition was symmetric, all
the codes develop asymmetries as the instabilities go non-linear.

In figure~\ref{fig:blobentropy} we show entropy slices in a zoomed-in
regions around the blob for several different methods on the test at
different times.  The top case is the standard {\sc ENZO} result.  The blob
test initially has uniform high entropy in the flow and low entropy in
the blob.  As the flow impacts the blob the shock creates even higher
entropy in the flow, shown in white in the figure.   The blob is
somewhat like a mushroom cloud on its side, where deceleration plays the role
of gravity and the high entropy gas
initially wraps around the exterior.  The shock rapidly fades and from
that point new entropy production is minimal.  In traditional SPH, the
lack of diffusion means that SPH particles retain their
entropy values for the rest of the run.   This is particularly
apparent in the second and fourth rows where no diffusion was used.
In particular, even with surface-tension removed (GDF), there are
dense, low entropy knots that move through the high low entropy
material.
Resolution-scale instabilities rapidly mix entropy on a local
crossing-time.  The precise numerical behaviour
depends on the absolute velocity in grid codes (three {\sc ENZO}
cases) and the mixing model (in SPH).  The primary difference between
the second last and final rows is the move to more neighbours (64 to 200) 
and the Wendland C4 kernel.  
The result is a reduction in noise and smoother features.   

\begin{figure*}
	\includegraphics[width=\textwidth]{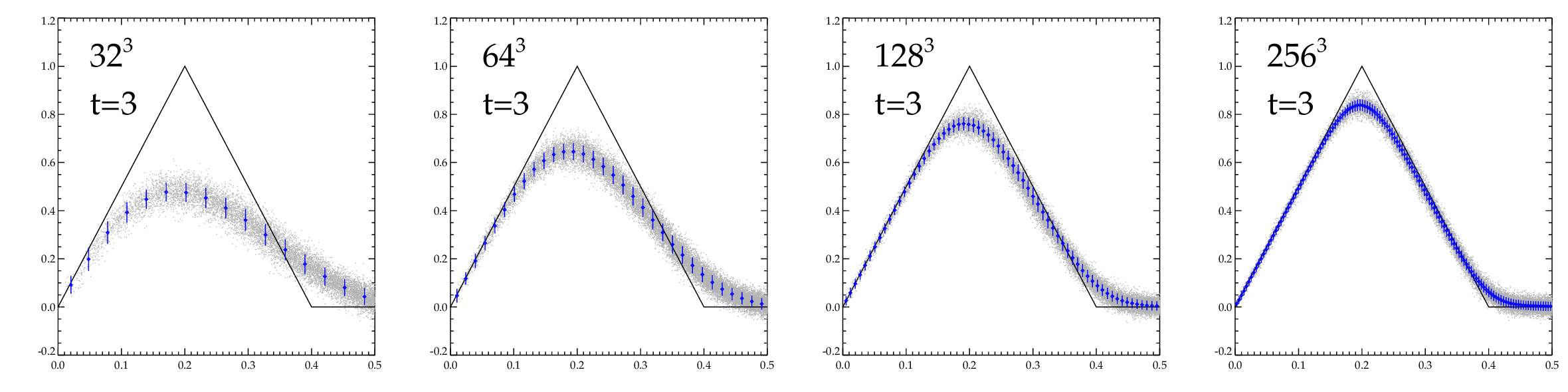}
    \caption{Velocity vs. radius in the Gresho-Chan Vortex with the full {\sc Gasoline2} method at t=3
      (2.4 rotations of the peak).  Resolution is indicated in the top left of each panel.  The blue
    bars are inserted at one particle separation as an indication of
    resolution and the thickness is an indicator of the {\it rms}
    deviation.  The gray points are individual particles.   The black
    curve is the exact solution. }
    \label{fig:gresho}
% Our gresho is full 3d -- who else did that?
\end{figure*}

%\begin{figure*}
%	\includegraphics[width=\textwidth]{greshoCDNa_nosc.ps}
%    \caption{{\sc Gasoline2} Gresho Vortex with $\alpha_{NOISE}$.} 
%    \label{fig:greshonoise}
% This figure is optional -- looks the same
%\end{figure*}

\begin{figure}
	\includegraphics[width=0.49\columnwidth]{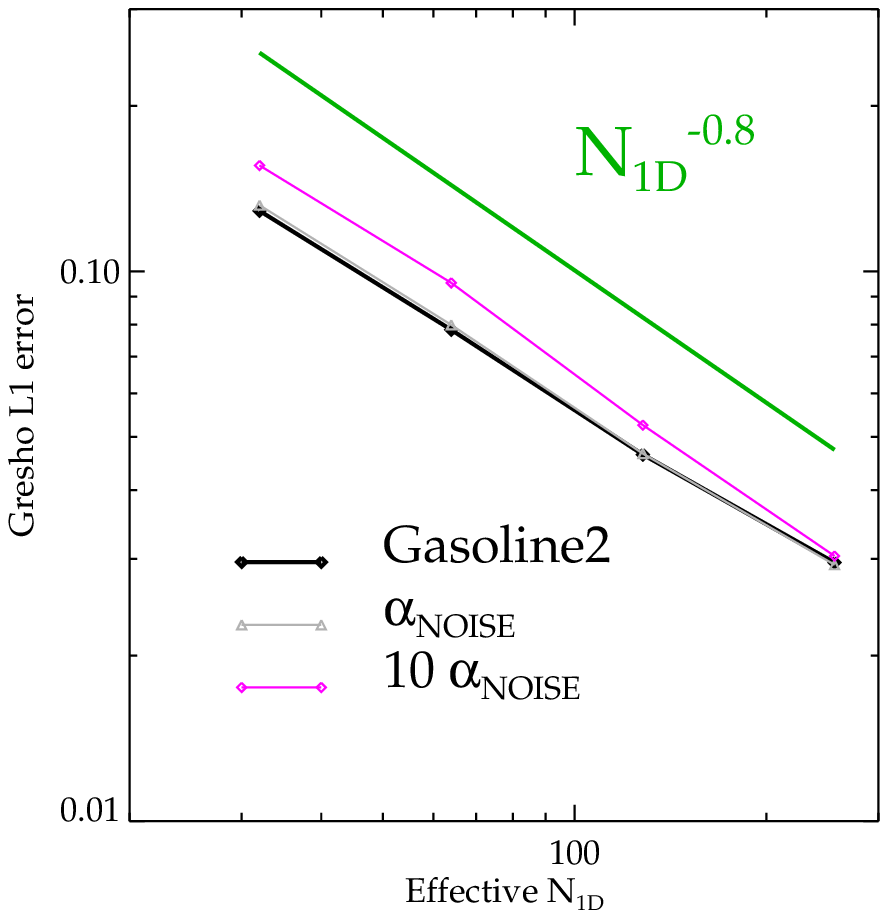}
	\includegraphics[width=0.49\columnwidth]{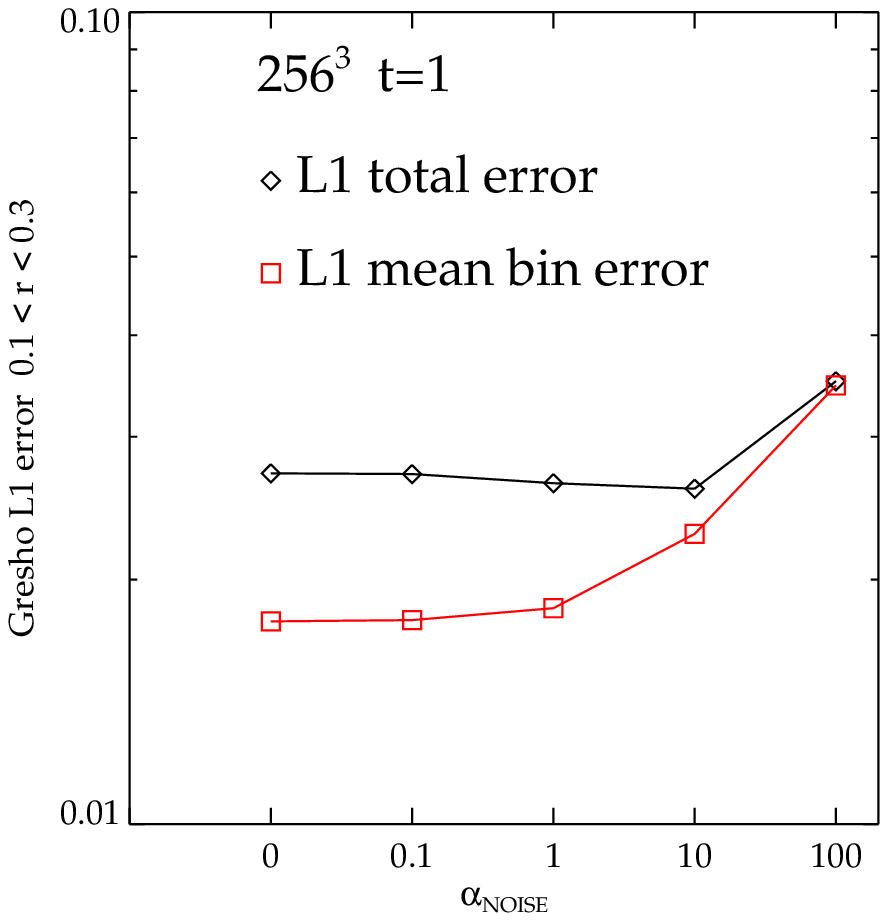}
    \caption{Gresho-Chan Vortex L1 errors.  Left: L1 error versus effective resolution for $N_{1D}$ = 32, 64, 128 and 256.  The convergence rate is roughly as $N_{1D}^{-0.8}$.  The Noise result is nearly on-top of the base result.  
Right: L1 errors versus amount of artificial viscosity introduced to combat noise for $0.1 < r < 0.3$ (near the peak of velocity).  The solid lines show the total L1 error and the dashed curves show the contribution to L1 error from the binned (mean) solution only.}
    \label{fig:greshoL1}
% Note: L1 is 50% particles completely outside the Vortex -- thus L1 mostly probes raw noise rather and barely includes errors at small r.
% This figure shows that noise suppression is tricky to get right c.f. Rosswog 2015's claim it gives better answers
% How I do noise is not described in text yet
\end{figure}

\subsection{Gresho-Chan Vortex}\label{tests:gresho}
The Gresho-Chan vortex \citep{Gresho1990} is a cylindrical vortex that is ideal for
testing how well a code can simulate an inviscid fluid.  The ideal
solution is steady, with centrifugal acceleration balancing pressure gradient.
Viscosity will cause angular momentum transport, disrupting this equilibrium by letting
the inner part of the vortex torque up the outer parts.  The vortex is
initialized with $\rho=1$, $-1<x,y<1$, and a piecewise pressure function,
%Gresho was 7-14 particles thick -- I tried both, no difference
\begin{equation}
    P(r) = 
    \begin{cases}
        5+12.5\,r^2 & \quad (0 \leq r < 0.2)\\
        9+12.5\,r^2-20\,r+4\ln{5r} & \quad (0.2 \leq r < 0.4)\\
        3+4\ln{2} & \quad (r \geq 0.4)\\
    \end{cases}
\end{equation}
and velocity function, $v_r=0$, and 
\begin{equation}
    v_\theta(r) = 
    \begin{cases}
        5r & \quad (0 \leq r < 0.2)\\
        2-5r & \quad (0.2 \leq r < 0.4)\\
        0 & \quad (r \geq 0.4).\\
    \end{cases}
\end{equation}
We evolve the vortex to $t=3$, or $\sim2.4$ rotations of the
peak.  We note in passing that not all code papers do this
\cite[e.g.][]{Hu2014}.  Given that the vortex decays in all cases, it
is mostly important for making fair comparisons.
A key choice was to evolve this problem using a glass in
three-dimensions,  where most others
\citep{Springel2010a,Kawata2013,Hopkins2015} evolve the problem in
two.   We refer the reader to \citet{Dehnen2012} for a detailed
discussion of this problem and how numerical factors affect it.

Figure~\ref{fig:gresho} shows the results at four different resolutions
at time $t=3$.  A viscosity limiter \citep[][CD]{Morris1997a} is
critical for this test.  The improvements present in the {\sc
  Gasoline2} limiter (section~\ref{sec:av}) do not change the outcome
from a more straightforward CD limiter on this problem.
 The convergence of the L1 norm error is fairly typical and
somewhat sub-linear ($N_{1D}^{-0.8}$) as shown in the left panel in
figure~\ref{fig:greshoL1}.

\cite{Rosswog2015} argued that triggering viscosity on noise could improve the
solution further.  We experimented with noise triggers similar to
those of \citet{Rosswog2015} as additions to the shock trigger of CD.   
For all the variants we found
that it was hard to achieve a substantial improvement over our standard scheme.
We include results of additional tests with a very simple noise trigger, 
$\alpha_{NOISE}=\sigma_v^2/(\sigma_v^2+c^2)$ where $\sigma_v$ is
the rms velocity noise at the particle to illustrate the general behaviour  (light line in the left panel in figure~\ref{fig:greshoL1}).     
Attempting strong noise suppression made the solution worse (diamond line, $10\,\alpha_{NOISE}$).
We attribute this partly to the fact that the shock trigger
also triggers on noise in the absence of strong expansion or shear to
suppress it so some noise viscosity is already present.   Examining
our results, we were also concerned that it was possible to reduce the
L1 error even though the mean solution (i.e. in radial bins) was worse.
This behaviour is demonstrated in the right panel in figure~\ref{fig:greshoL1} 
which
measures the error near the peak of the vortex.   We chose the peak
because there are many particles outside $r=0.3$ that otherwise
dominate the measurement.
The L1 norm is a combination of the spread and the mean deviation from the
exact solution.  We find that extra viscosity always makes the mean
solution worse (squares in the right panel in fig.~\ref{fig:greshoL1}) 
even if the
overall L1 error is modestly improved.    The velocity distributions with the noise trigger are
visually indistinguishable from the results shown in figure~\ref{fig:gresho}.
For now the standard {\sc   Gasoline2} has no noise trigger but a careful re-examination might yield
benefits from employing one.

\begin{figure}
	\includegraphics[width=\columnwidth]{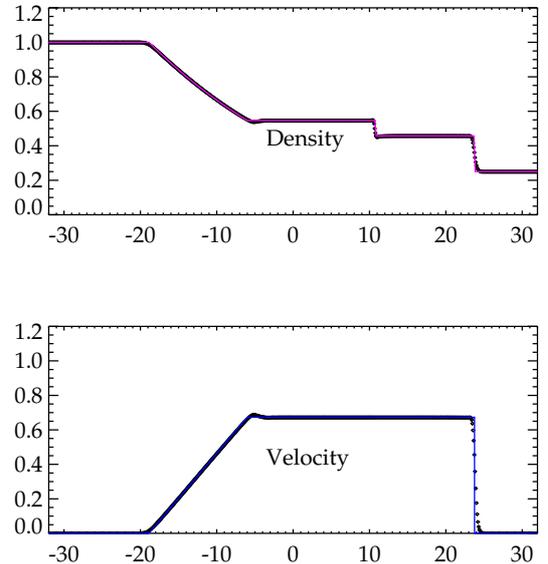}
    \caption{Sod Shock Tube for {\sc Gasoline2} without variable
      viscosity.  The symbols show averages values
      separated by the local 1d particle spacing to indicate the resolution and the
      thin lines indicate the exact solution.
      The top panel shows
    density and the lower is velocity.}
    \label{fig:sod}
\end{figure}

\begin{figure}
	\includegraphics[width=\columnwidth]{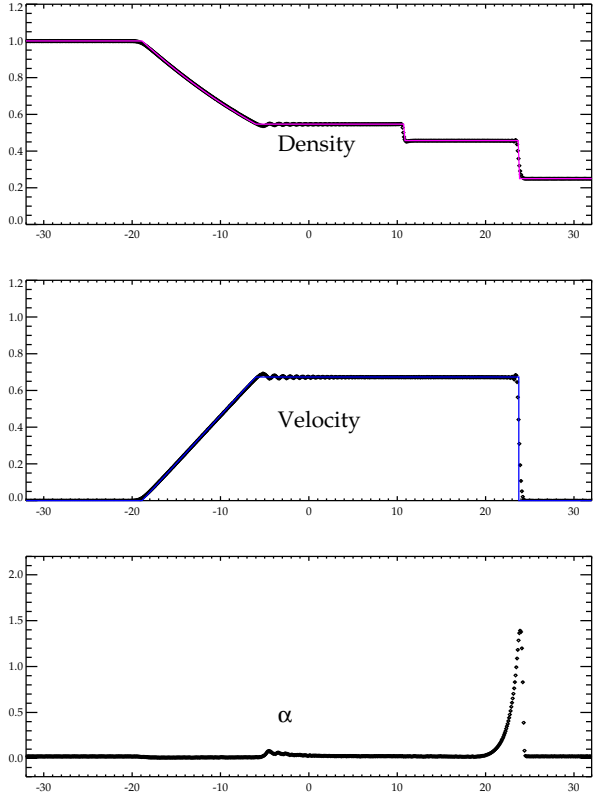}
    \caption{Same Sod Shock Tube as figure~\ref{fig:sod} for {\sc Gasoline2}
      with the viscosity limiter.  The additional lower panel shows
      the local viscosity $\alpha$.}
    \label{fig:sodcd}
% Shows no free lunch -- reducing alpha from 1 away from shows allows ringing 
% Note ringing can be seen in PSPH solution in gizmo paper
\end{figure}

\subsection{Sod Shock Tube}\label{tests:sod}
No code paper would  be complete without the classic Sod shock tube
\citep{Sod1978}.  This simple 1D Riemann problem begins with two domains at rest,
but out of pressure equilibrium.  The initial conditions are $\rho_{left}=1$, $P_{left}=1$ for $x<0$ 
and $\rho_{right}=0.25$, $P_{right}=0.1795$ for $x>0$.
The self-similar solution consists of a rarefaction fan traveling left, a contact discontinuity, 
and shock moving to the right and is shown in the figures as a black line.

Figure~\ref{fig:sod} shows {\sc Gasoline2} without variable viscosity but
with the Balsara limiter.
This is comparable to standard {\sc Gasoline} as of 2011 with the addition of the Wendland
C4 kernel and 200 neighbours, $\alpha=2$ and $\beta=4$.  The use of
200 neighbours and the large shock parameters increases the shock
width over versions with older kernels, 50-64 neighbours, $\alpha=1$
and $\beta=2$.  Choosing $\alpha=1$ substantially decreases the shock
width and allows some modest post-shock ringing as seen in
\citet{Wadsley2004}.

Figure~\ref{fig:sodcd} shows the full {\sc Gasoline2} result, including the
viscosity limiter on same initial conditions as figure~\ref{fig:sod}.
The solution left of the initial contact has very little viscosity and
experiences mild ringing at the velocity peak of the rarefaction wave.
All variants of the time-dependent viscosity suffer in the
rarefaction wave where there is no viscosity until after it peaks.  Applying
a minimum alpha was not found to be helpful in reducing this unless it
was large enough to negatively affect other tests (e.g. Gresho-Chan
Vortex).
The method supplies an $\alpha \sim 0.1-0.2$ in the post-rarefaction region, as was also seen
by CD on this test.   We note that the velocity in this region for PSPH also shows some
mild ringing \citep{Hopkins2015,Hu2014}.
% Test if it is the 200 neighbours at fault here?
% would a minimum alpha help?

\begin{figure*}
	\includegraphics[width=\textwidth]{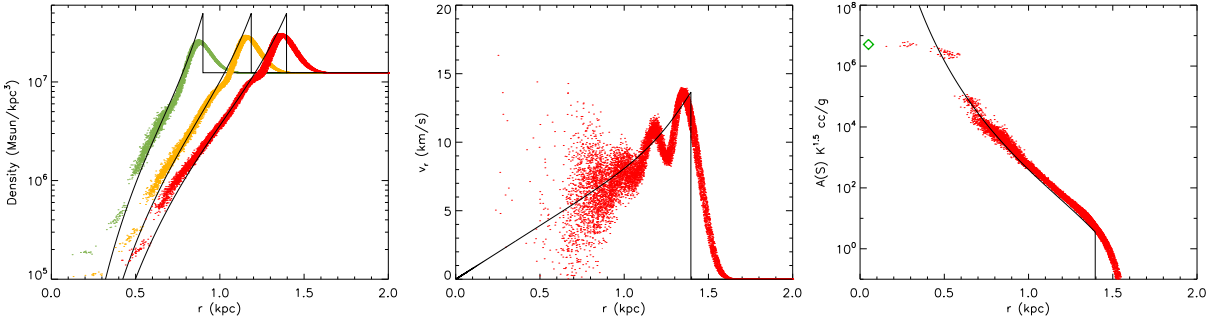}
    \caption{Sedov-Taylor explosion results for zero external temperatures.  This is both an extremely strong shock and a test for accurate time integration.  The panels show density at three times (left), velocity at the final time (centre) and entropic function at the final time (right).  Note that the exact solution (black lines) is for a point explosion. }
    \label{fig:sedov}
%Note ringing shown in Hu+ paper velocity profile
%Note velocity exact profile line is correct
% Based on astrosim IC -- 64 hot particles
\end{figure*}

\subsection{Sedov-Taylor Blast}\label{tests:sedov}

The Sedov-Taylor blast wave \citep{Taylor1950,Sedov1959} is a spherical shockwave
generated by the injection of energy in a central region.  For this test, we
have a $128^3$ box, with 64 particles injected with $6.78\times10^{53}\;\rm{erg}$.
The background fluid has temperature, $T=0$, making this an infinite Mach number shock,
with a density enhancement of 4 for a $\gamma=5/3$ equation of state.  The
initial conditions have a domain of $6\times6\times6\;\rm{kpc}$, and a density
of $0.5\;\rm{cm^{-3}}$.   This test will break codes without careful time step adjustment as
demonstrated in \citet{Saitoh2009}.    The total energy error between the
start and the final time in the {\sc Gasoline2} run was $0.2\%.$ %Wc4_n200_nosc

Figure~\ref{fig:sedov} shows results for {\sc Gasoline2} at three
times.  For the final time we also show the velocity and entropy
function profiles.  The initial condition had a finite initial hot
mass with a maximum entropy indicated by the diamond in the right
panel so the entropy follows the exact solution until it gets to this
value.   In the left panel we plot the density on log axes at three
equally spaced times.  The solution deviates near the point where the
entropy limit kicks in.  The central panel shows velocity which is
commonly left out in plots
of these results.   As shown in \citet{Hu2014} we see considerable
ringing in the velocity.  We have verified that the viscosity rises to
$\alpha=4$,
its maximum value, well before the shock.   The perfectly cold
(T=0) pre-shock gas essentially forces this.   This is an infinite
Mach number shock and thus a strenuous test of the code.  We have
experimented with larger $\alpha_{max}$ values but they both broaden
the shock and negatively affect other aspects of the solution for this
and other tests.  Thus we show the current results as an acceptable compromise.
% Velocity ringing is ugly, not everyone shows it.  Not sure how much we can do AlphaMax already = 4  Note that Alpha is on strongly immediately too (all r > r_shock) due to c=0 there.

\begin{figure}
	\includegraphics[width=\columnwidth]{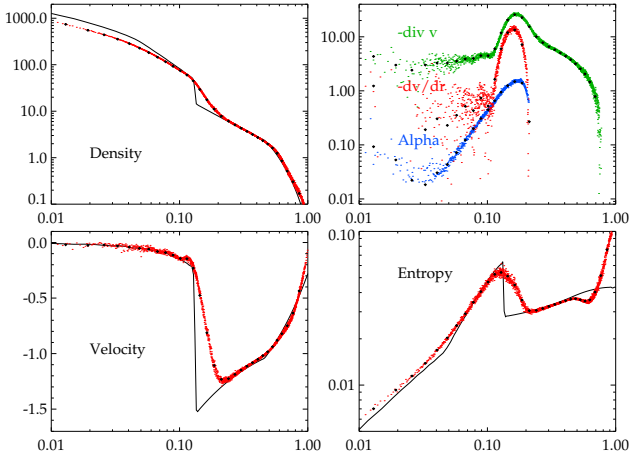}
    \caption{Evrard adiabatic collapse test at time $t=0.8$ with
      standard {\sc Gasoline2} parameters.   The four panels show
      Density, Velocity, Entropy and properties relevant to shock
      capturing.  Note that Entropy does not increase until the
      particles are within range of the actual shock near $r=0.1$.}
    \label{fig:evrardwc4n200}
\end{figure}

\begin{figure}
	\includegraphics[width=\columnwidth]{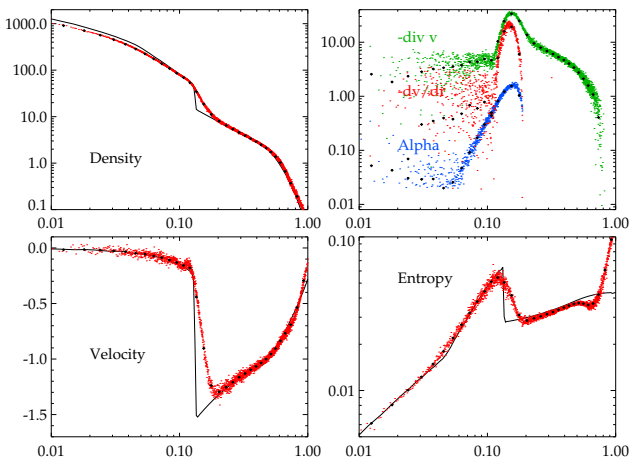}
    \caption{Evrard adiabatic collapse test at time $t=0.8$ except
      with the Wendland C2 kernel and 64 neighbours.}
    \label{fig:evrardw64}
% Demonstrates no free lunch.  While some problems that depend on accurate velocity fields (e.g. gresho) benefit
% from more nbrs (less E0), for strong shocks features are broadened and errors due to E0 are minor/irrelevant
\end{figure}

\subsection{Evrard Collapse Test}\label{test:evrard}

The adiabatic gas-only collapse test from \cite{Evrard1988} includes a
strong shock and cold pre-shock infall.  It is the only test shown here that requires gravity.
It can be used to assess the ability of a code to follow the 
collapse phase during the formation of an astrophysical object.  
It tests shock capturing and the coupling between gravity and gas
dynamics.   We use a glass initial condition with 28000 particles,
comparable to the resolution in the original papers.    The error in
total energy over the course of the full {\sc Gasoline2} run was $0.04\%$

Figure~\ref{fig:evrardwc4n200} shows the structure at at time $t=0.8$
($t=0.88$ in Hernquist \& Katz, 1989).
The results shown as diamonds are binned at the particle spacing with actual
particle values shown as lighter points.  The solid line is a high
resolution 1D PPM solution from \cite{Steinmetz1993}.
Prior SPH results suffer from pre-shock entropy production as the compression triggers
the artificial viscosity.  In SPH codes that we are aware of (including {\sc Gasoline} in
2004), roughly half the entropy is produced during the pre-shock
infall because viscosity is triggered on negative divergence.

Using a {\it gradient based} shock indicator based on $\frac{dv}{dn}$ rather than
divergence we detect the shock at the right place.  The extra shock
width is solely due to the finite resolution.  The difference is highlighted in the
top right panel of figure~\ref{fig:evrardwc4n200}, where we show that
our shock viscosity $\alpha$ turns on at the right place, where
$\frac{dv}{dr}$ switches sign from positive to negative.   As
discussed in section~\ref{sec:av}, in a collapse like this the divergence
is dominated by the geometric contraction of the solid angle rather
than the radial velocity gradient.  Using gradient-based shock
detection works well in this and any geometry.  It also filters out
uniform contraction or expansion.

Figure~\ref{fig:evrardw64} shows {\sc Gasoline2} results with less
conservative smoothing (Wendland C2 kernel and 64 neighbours).  This
improves the resolution (central density and narrower
shock) at the cost of increased noise everywhere, particularly in
velocity.   This illustrates that in large scale collapse problems
with orders of magnitude increases in density, the benefits of smaller
local errors associated with large neighbour counts are less obvious.

\section{Conclusions}\label{concl}

This paper provides a complete description of the SPH methods present in
{\sc Gasoline2}.  We argue that these improvements or similar ones are
required for any modern SPH code to overcome the limitations of
traditional SPH.  In the following summary we have highlighted methods
unique to {\sc Gasoline2} with italics:
\begin{itemize}
    \item {\it Geometric Density Average Force} and internal energy
        expressions which both
        minimize surface tension effects in multiphase flow and
        naturally provide excellent entropy conservation.
    \item {\it Turbulent Diffusion}, first introduced in \citet{Wadsley2008},
        which models physically necessary sub-grid turbulent mixing, alleviating 
        problems with hydrodynamic instabilities and scalar transport
        (e.g. entropy).
    \item A {\it Gradient-Based} shock detector based on the
      one-dimensional velocity gradient normal to shocks,
      rather than divergence.  This prevents spurious viscosity in 
      convergent flows such as spherical collapse problems.
    \item Time-dependent artificial viscosity, modified from the prescription of
        \citet{Morris1997a} and \citet{Cullen2010}, to reduce velocity noise and
        better treat inviscid flow.
    \item Wendland kernels, shown by \citet{Dehnen2012} to allow larger
        neighbour numbers without pairing and improve convergence.
    \item Time step adjustment, shown by \citet{Saitoh2009} to be necessary for
        handling sudden jumps in time step such as due to rapid heating
        in strong shocks.
\end{itemize}

We have presented test problems that show the necessity of each of
these components to give statisfactory results.  
The tests presented here cover many situations relevant to
astrophysical simulations.  Thus {\sc Gasoline2} and {\sc Changa}
should be regarded as cutting-edge, modern SPH codes.  
It is worth emphasizing
that we use a Density-Energy SPH method.  Pressure-Entropy formulations
are not required prevent surface tension problems.   In fact, the distinction between
these method has been blurred via the use of many different density weightings 
\citep[e.g.][]{Rosswog2015}.

Lagrangian techniques such as SPH offer considerable advantages for
astrophysical simulations (e.g. natural adaptivity, Galilean
invariance, accurate orbits and efficient time stepping).  We note
that SPH is still somewhat more diffusive and slower to converge than
other methods for a given resolution.  New methods have
arisen that offer faster convergence through the use of more complex
gradient estimates and fluxes from Riemann solvers, such as AREPO \citep{Springel2010a} and
GIZMO \citep{Hopkins2015}.  AREPO suffered from some SPH-like
numerical convergence issues which have now been resolved \citep{Mocz2015,Pakmor2016}. New SPH
methods with accurate gradients based on integral form (ISPH) also
show considerable promise  without the cost of using Riemann solvers
\citep{Cabezon2016}.  \cite{Valdarnini2016}
showed promising ISPH results for previously problematic regimes such
as subsonic turbulence.   Thus
continuous code improvement and testing new methods seems to be the lot of the
computational astrophysicist for the foreseeable future.

\section*{Acknowledgements}

The authors would like to thank Hugh Couchman, Walter Dehnen, Jonathan Panuelos
and Zachariah Levine for helpful interactions during the preparation of this
manuscript.  We would also like the thank the referee, Daniel Price, for helpful suggestions.
The authors would also like to thank SHARCNET/Compute Canada for
HPC resources and support and the Natural Sciences and Engineering Research
Council of Canada (NSERC) for financial support. Tom Quinn acknowledges support
from NSF award AST-1311956.

%%%%%%%%%%%%%%%%%%%%%%%%%%%%%%%%%%%%%%%%%%%%%%%%%%

%%%%%%%%%%%%%%%%%%%% REFERENCES %%%%%%%%%%%%%%%%%%

\bibliographystyle{mnras}
\bibliography{references} 

\begin{thebibliography}{}
\makeatletter
\relax
\def\mn@urlcharsother{\let\do\@makeother \do\$\do\&\do\#\do\^\do\_\do\%\do\~}
\def\mn@doi{\begingroup\mn@urlcharsother \@ifnextchar [ {\mn@doi@}
  {\mn@doi@[]}}
\def\mn@doi@[#1]#2{\def\@tempa{#1}\ifx\@tempa\@empty \href
  {http://dx.doi.org/#2} {doi:#2}\else \href {http://dx.doi.org/#2} {#1}\fi
  \endgroup}
\def\mn@eprint#1#2{\mn@eprint@#1:#2::\@nil}
\def\mn@eprint@arXiv#1{\href {http://arxiv.org/abs/#1} {{\tt arXiv:#1}}}
\def\mn@eprint@dblp#1{\href {http://dblp.uni-trier.de/rec/bibtex/#1.xml}
  {dblp:#1}}
\def\mn@eprint@#1:#2:#3:#4\@nil{\def\@tempa {#1}\def\@tempb {#2}\def\@tempc
  {#3}\ifx \@tempc \@empty \let \@tempc \@tempb \let \@tempb \@tempa \fi \ifx
  \@tempb \@empty \def\@tempb {arXiv}\fi \@ifundefined
  {mn@eprint@\@tempb}{\@tempb:\@tempc}{\expandafter \expandafter \csname
  mn@eprint@\@tempb\endcsname \expandafter{\@tempc}}}

\bibitem[\protect\citeauthoryear{{Agertz} et~al.,}{{Agertz}
  et~al.}{2007}]{Agertz2007}
{Agertz} O.,  et~al., 2007, \mn@doi [\mnras]
  {10.1111/j.1365-2966.2007.12183.x}, \href
  {http://adsabs.harvard.edu/abs/2007MNRAS.380..963A} {380, 963}

\bibitem[\protect\citeauthoryear{{Balsara}}{{Balsara}}{1995}]{Balsara1995}
{Balsara} D.~S.,  1995, \mn@doi [Journal of Computational Physics]
  {10.1016/S0021-9991(95)90221-X}, \href
  {http://adsabs.harvard.edu/abs/1995JCoPh.121..357B} {121, 357}

\bibitem[\protect\citeauthoryear{{Bauer} \& {Springel}}{{Bauer} \&
  {Springel}}{2012}]{Bauer2012}
{Bauer} A.,  {Springel} V.,  2012, \mn@doi [\mnras]
  {10.1111/j.1365-2966.2012.21058.x}, \href
  {http://adsabs.harvard.edu/abs/2012MNRAS.423.2558B} {423, 2558}

\bibitem[\protect\citeauthoryear{{Beck} et~al.,}{{Beck}
  et~al.}{2016}]{Beck2016}
{Beck} A.~M.,  et~al., 2016, \mn@doi [\mnras] {10.1093/mnras/stv2443}, \href
  {http://adsabs.harvard.edu/abs/2016MNRAS.455.2110B} {455, 2110}

\bibitem[\protect\citeauthoryear{{Benincasa}, {Tasker}, {Pudritz}  \&
  {Wadsley}}{{Benincasa} et~al.}{2013}]{Benincasa2013}
{Benincasa} S.~M.,  {Tasker} E.~J.,  {Pudritz} R.~E.,   {Wadsley} J.,  2013,
  \mn@doi [\apj] {10.1088/0004-637X/776/1/23}, \href
  {http://adsabs.harvard.edu/abs/2013ApJ...776...23B} {776, 23}

\bibitem[\protect\citeauthoryear{Benz}{Benz}{1990}]{Benz1990}
Benz W.,  1990, Smooth Particle Hydrodynamics: A Review.
Springer Netherlands, Dordrecht, pp 269--288,
  \mn@doi{10.1007/978-94-009-0519-1_16}, \url
  {http://dx.doi.org/10.1007/978-94-009-0519-1_16}

\bibitem[\protect\citeauthoryear{{Bryan} \& {Norman}}{{Bryan} \&
  {Norman}}{1997}]{Bryan1997}
{Bryan} G.~L.,  {Norman} M.~L.,  1997, in {Clarke} D.~A.,  {West} M.~J.,  eds,
  Astronomical Society of the Pacific Conference Series Vol. 123, Computational
  Astrophysics; 12th Kingston Meeting on Theoretical Astrophysics. p.~363
  (\mn@eprint {} {astro-ph/9710186})

\bibitem[\protect\citeauthoryear{{Bryan} et~al.,}{{Bryan}
  et~al.}{2014}]{Bryan2014}
{Bryan} G.~L.,  et~al., 2014, \mn@doi [\apjs] {10.1088/0067-0049/211/2/19},
  \href {http://adsabs.harvard.edu/abs/2014ApJS..211...19B} {211, 19}

\bibitem[\protect\citeauthoryear{{Cabezon}, {Garcia-Senz}  \&
  {Figueira}}{{Cabezon} et~al.}{2016}]{Cabezon2016}
{Cabezon} R.~M.,  {Garcia-Senz} D.,   {Figueira} J.,  2016, preprint, \href
  {http://adsabs.harvard.edu/abs/2016arXiv160701698C} {} (\mn@eprint {arXiv}
  {1607.01698})

\bibitem[\protect\citeauthoryear{{Cullen} \& {Dehnen}}{{Cullen} \&
  {Dehnen}}{2010}]{Cullen2010}
{Cullen} L.,  {Dehnen} W.,  2010, \mn@doi [\mnras]
  {10.1111/j.1365-2966.2010.17158.x}, \href
  {http://adsabs.harvard.edu/abs/2010MNRAS.408..669C} {408, 669}

\bibitem[\protect\citeauthoryear{Dehnen}{Dehnen}{2016}]{Dehnen2016}
Dehnen W.,  2016, Private Communication

\bibitem[\protect\citeauthoryear{{Dehnen} \& {Aly}}{{Dehnen} \&
  {Aly}}{2012}]{Dehnen2012}
{Dehnen} W.,  {Aly} H.,  2012, \mn@doi [\mnras]
  {10.1111/j.1365-2966.2012.21439.x}, \href
  {http://adsabs.harvard.edu/abs/2012MNRAS.425.1068D} {425, 1068}

\bibitem[\protect\citeauthoryear{{Evrard}}{{Evrard}}{1988}]{Evrard1988}
{Evrard} A.~E.,  1988, \mn@doi [\mnras] {10.1093/mnras/235.3.911}, \href
  {http://adsabs.harvard.edu/abs/1988MNRAS.235..911E} {235, 911}

\bibitem[\protect\citeauthoryear{{Frenk} et~al.,}{{Frenk}
  et~al.}{1999}]{Frenk1999}
{Frenk} C.~S.,  et~al., 1999, \mn@doi [\apj] {10.1086/307908}, \href
  {http://adsabs.harvard.edu/abs/1999ApJ...525..554F} {525, 554}

\bibitem[\protect\citeauthoryear{{Fryxell} et~al.,}{{Fryxell}
  et~al.}{2000}]{Fryxell2000}
{Fryxell} B.,  et~al., 2000, \mn@doi [\apjs] {10.1086/317361}, \href
  {http://adsabs.harvard.edu/abs/2000ApJS..131..273F} {131, 273}

\bibitem[\protect\citeauthoryear{{Gaburov} \& {Nitadori}}{{Gaburov} \&
  {Nitadori}}{2011}]{Gaburov2011}
{Gaburov} E.,  {Nitadori} K.,  2011, \mn@doi [\mnras]
  {10.1111/j.1365-2966.2011.18313.x}, \href
  {http://adsabs.harvard.edu/abs/2011MNRAS.414..129G} {414, 129}

\bibitem[\protect\citeauthoryear{{Germano}, {Piomelli}, {Moin}  \&
  {Cabot}}{{Germano} et~al.}{1991}]{Germano1991}
{Germano} M.,  {Piomelli} U.,  {Moin} P.,   {Cabot} W.~H.,  1991, \mn@doi
  [Physics of Fluids] {10.1063/1.857955}, \href
  {http://adsabs.harvard.edu/abs/1991PhFl....3.1760G} {3, 1760}

\bibitem[\protect\citeauthoryear{{Gibson}, {Courty},
  {S{\'a}nchez-Bl{\'a}zquez}, {Teyssier}, {House}, {Brook}  \&
  {Kawata}}{{Gibson} et~al.}{2009}]{Gibson2009}
{Gibson} B.~K.,  {Courty} S.,  {S{\'a}nchez-Bl{\'a}zquez} P.,  {Teyssier} R.,
  {House} E.~L.,  {Brook} C.~B.,   {Kawata} D.,  2009, in {Andersen} J.,
  {Nordstr{\"o}ara} {m} B.,   {Bland-Hawthorn} J.,  eds,  IAU Symposium Vol.
  254, The Galaxy Disk in Cosmological Context. pp 445--452 (\mn@eprint {arXiv}
  {0808.0576}), \mn@doi{10.1017/S1743921308027956}

\bibitem[\protect\citeauthoryear{{Gingold} \& {Monaghan}}{{Gingold} \&
  {Monaghan}}{1977}]{Gingold1977}
{Gingold} R.~A.,  {Monaghan} J.~J.,  1977, \mn@doi [\mnras]
  {10.1093/mnras/181.3.375}, \href
  {http://adsabs.harvard.edu/abs/1977MNRAS.181..375G} {181, 375}

\bibitem[\protect\citeauthoryear{Godunov}{Godunov}{1959}]{Godunov1959}
Godunov S.~K.,  1959, Matematicheskii Sbornik, 89, 271

\bibitem[\protect\citeauthoryear{{Gresho} \& {Chan}}{{Gresho} \&
  {Chan}}{1990}]{Gresho1990}
{Gresho} P.~M.,  {Chan} S.~T.,  1990, \mn@doi [International Journal for
  Numerical Methods in Fluids] {10.1002/fld.1650110510}, \href
  {http://adsabs.harvard.edu/abs/1990IJNMF..11..621G} {11, 621}

\bibitem[\protect\citeauthoryear{{Hernquist} \& {Katz}}{{Hernquist} \&
  {Katz}}{1989}]{Hernquist1989}
{Hernquist} L.,  {Katz} N.,  1989, \mn@doi [\apjs] {10.1086/191344}, \href
  {http://adsabs.harvard.edu/abs/1989ApJS...70..419H} {70, 419}

\bibitem[\protect\citeauthoryear{{Hopkins}}{{Hopkins}}{2013}]{Hopkins2013}
{Hopkins} P.~F.,  2013, \mn@doi [\mnras] {10.1093/mnras/sts210}, \href
  {http://adsabs.harvard.edu/abs/2013MNRAS.428.2840H} {428, 2840}

\bibitem[\protect\citeauthoryear{{Hopkins}}{{Hopkins}}{2015}]{Hopkins2015}
{Hopkins} P.~F.,  2015, \mn@doi [\mnras] {10.1093/mnras/stv195}, \href
  {http://adsabs.harvard.edu/abs/2015MNRAS.450...53H} {450, 53}

\bibitem[\protect\citeauthoryear{{Hu}, {Naab}, {Walch}, {Moster}  \&
  {Oser}}{{Hu} et~al.}{2014}]{Hu2014}
{Hu} C.-Y.,  {Naab} T.,  {Walch} S.,  {Moster} B.~P.,   {Oser} L.,  2014,
  preprint, \href {http://adsabs.harvard.edu/abs/2014arXiv1402.1788H} {}
  (\mn@eprint {arXiv} {1402.1788})

\bibitem[\protect\citeauthoryear{Kale \& Krishnan}{Kale \&
  Krishnan}{1993}]{Kale1993}
Kale L.~V.,  Krishnan S.,  1993, in Proceedings of the Eighth Annual Conference
  on Object-oriented Programming Systems, Languages, and Applications. OOPSLA
  '93.
ACM, New York, NY, USA, pp 91--108, \mn@doi{10.1145/165854.165874}, \url
  {http://doi.acm.org/10.1145/165854.165874}

\bibitem[\protect\citeauthoryear{{Kawata}, {Okamoto}, {Gibson}, {Barnes}  \&
  {Cen}}{{Kawata} et~al.}{2013}]{Kawata2013}
{Kawata} D.,  {Okamoto} T.,  {Gibson} B.~K.,  {Barnes} D.~J.,   {Cen} R.,
  2013, \mn@doi [\mnras] {10.1093/mnras/sts161}, \href
  {http://adsabs.harvard.edu/abs/2013MNRAS.428.1968K} {428, 1968}

\bibitem[\protect\citeauthoryear{{Keller}, {Wadsley}, {Benincasa}  \&
  {Couchman}}{{Keller} et~al.}{2014}]{Keller2014}
{Keller} B.~W.,  {Wadsley} J.,  {Benincasa} S.~M.,   {Couchman} H.~M.~P.,
  2014, \mn@doi [\mnras] {10.1093/mnras/stu1058}, \href
  {http://adsabs.harvard.edu/abs/2014MNRAS.442.3013K} {442, 3013}

\bibitem[\protect\citeauthoryear{{Kim} et~al.,}{{Kim} et~al.}{2014}]{Kim2014}
{Kim} J.-h.,  et~al., 2014, \mn@doi [\apjs] {10.1088/0067-0049/210/1/14}, \href
  {http://adsabs.harvard.edu/abs/2014ApJS..210...14K} {210, 14}

\bibitem[\protect\citeauthoryear{{Kitsionas} et~al.,}{{Kitsionas}
  et~al.}{2009}]{Kitsionas2009}
{Kitsionas} S.,  et~al., 2009, \mn@doi [\aap] {10.1051/0004-6361/200811170},
  \href {http://adsabs.harvard.edu/abs/2009A%26A...508..541K} {508, 541}

\bibitem[\protect\citeauthoryear{Kurganov \& Tadmor}{Kurganov \&
  Tadmor}{2000}]{Kurganov2000}
Kurganov A.,  Tadmor E.,  2000, \mn@doi [Journal of Computational Physics]
  {http://dx.doi.org/10.1006/jcph.2000.6459}, 160, 241

\bibitem[\protect\citeauthoryear{{Lucy}}{{Lucy}}{1977}]{Lucy1977}
{Lucy} L.~B.,  1977, \mn@doi [\aj] {10.1086/112164}, \href
  {http://adsabs.harvard.edu/abs/1977AJ.....82.1013L} {82, 1013}

\bibitem[\protect\citeauthoryear{{McNally}, {Lyra}  \& {Passy}}{{McNally}
  et~al.}{2012}]{McNally2012}
{McNally} C.~P.,  {Lyra} W.,   {Passy} J.-C.,  2012, \mn@doi [\apjs]
  {10.1088/0067-0049/201/2/18}, \href
  {http://adsabs.harvard.edu/abs/2012ApJS..201...18M} {201, 18}

\bibitem[\protect\citeauthoryear{{Menon}, {Wesolowski}, {Zheng}, {Jetley},
  {Kale}, {Quinn}  \& {Governato}}{{Menon} et~al.}{2015}]{Menon2015}
{Menon} H.,  {Wesolowski} L.,  {Zheng} G.,  {Jetley} P.,  {Kale} L.,  {Quinn}
  T.,   {Governato} F.,  2015, \mn@doi [Computational Astrophysics and
  Cosmology] {10.1186/s40668-015-0007-9}, \href
  {http://adsabs.harvard.edu/abs/2015ComAC...2....1M} {2, 1}

\bibitem[\protect\citeauthoryear{{Mocz}, {Vogelsberger}, {Pakmor}, {Genel},
  {Springel}  \& {Hernquist}}{{Mocz} et~al.}{2015}]{Mocz2015}
{Mocz} P.,  {Vogelsberger} M.,  {Pakmor} R.,  {Genel} S.,  {Springel} V.,
  {Hernquist} L.,  2015, \mn@doi [\mnras] {10.1093/mnras/stv1598}, \href
  {http://adsabs.harvard.edu/abs/2015MNRAS.452.3853M} {452, 3853}

\bibitem[\protect\citeauthoryear{Monaghan}{Monaghan}{1987}]{Monaghan1987}
Monaghan J.,  1987, Monash University technical report, SPH Meets the Shocks of
  Noh

\bibitem[\protect\citeauthoryear{{Monaghan}}{{Monaghan}}{1992}]{Monaghan1992}
{Monaghan} J.~J.,  1992, \mn@doi [\araa] {10.1146/annurev.aa.30.090192.002551},
  \href {http://adsabs.harvard.edu/abs/1992ARA%26A..30..543M} {30, 543}

\bibitem[\protect\citeauthoryear{{Monaghan}}{{Monaghan}}{1997}]{Morris1997b}
{Monaghan} J.~J.,  1997, \mn@doi [Journal of Computational Physics]
  {10.1006/jcph.1997.5732}, \href
  {http://adsabs.harvard.edu/abs/1997JCoPh.136..298M} {136, 298}

\bibitem[\protect\citeauthoryear{{Monaghan}}{{Monaghan}}{2005}]{Monaghan2005}
{Monaghan} J.~J.,  2005, \mn@doi [Reports on Progress in Physics]
  {10.1088/0034-4885/68/8/R01}, \href
  {http://adsabs.harvard.edu/abs/2005RPPh...68.1703M} {68, 1703}

\bibitem[\protect\citeauthoryear{Morris \& Monaghan}{Morris \&
  Monaghan}{1997}]{Morris1997a}
Morris J.,  Monaghan J.,  1997, \mn@doi [Journal of Computational Physics]
  {http://dx.doi.org/10.1006/jcph.1997.5690}, 136, 41

\bibitem[\protect\citeauthoryear{Mott \& Oran}{Mott \& Oran}{2001}]{Mott2001}
Mott D.~R.,  Oran E.~S.,  2001, Technical report, CHEMEQ2: A solver for the
  stiff ordinary differential equations of chemical kinetics.
DTIC Document

\bibitem[\protect\citeauthoryear{{Nelson} \& {Papaloizou}}{{Nelson} \&
  {Papaloizou}}{1994}]{Nelson1994}
{Nelson} R.~P.,  {Papaloizou} J.~C.~B.,  1994, \mn@doi [\mnras]
  {10.1093/mnras/270.1.1}, \href
  {http://adsabs.harvard.edu/abs/1994MNRAS.270....1N} {270, 1}

\bibitem[\protect\citeauthoryear{{O'Shea}, {Nagamine}, {Springel}, {Hernquist}
  \& {Norman}}{{O'Shea} et~al.}{2005}]{OShea2005}
{O'Shea} B.~W.,  {Nagamine} K.,  {Springel} V.,  {Hernquist} L.,   {Norman}
  M.~L.,  2005, \mn@doi [\apjs] {10.1086/432645}, \href
  {http://adsabs.harvard.edu/abs/2005ApJS..160....1O} {160, 1}

\bibitem[\protect\citeauthoryear{{Pakmor}, {Springel}, {Bauer}, {Mocz},
  {Munoz}, {Ohlmann}, {Schaal}  \& {Zhu}}{{Pakmor} et~al.}{2016}]{Pakmor2016}
{Pakmor} R.,  {Springel} V.,  {Bauer} A.,  {Mocz} P.,  {Munoz} D.~J.,
  {Ohlmann} S.~T.,  {Schaal} K.,   {Zhu} C.,  2016, \mn@doi [\mnras]
  {10.1093/mnras/stv2380}, \href
  {http://adsabs.harvard.edu/abs/2016MNRAS.455.1134P} {455, 1134}

\bibitem[\protect\citeauthoryear{{Potter}, {Stadel}  \& {Teyssier}}{{Potter}
  et~al.}{2016}]{Potter2016}
{Potter} D.,  {Stadel} J.,   {Teyssier} R.,  2016, preprint, \href
  {http://adsabs.harvard.edu/abs/2016arXiv160908621P} {} (\mn@eprint {arXiv}
  {1609.08621})

\bibitem[\protect\citeauthoryear{{Press}, {Teukolsky}, {Vetterling}  \&
  {Flannery}}{{Press} et~al.}{1992}]{Press1992}
{Press} W.~H.,  {Teukolsky} S.~A.,  {Vetterling} W.~T.,   {Flannery} B.~P.,
  1992, {Numerical recipes in C. The art of scientific computing}

\bibitem[\protect\citeauthoryear{{Price}}{{Price}}{2008}]{Price2008}
{Price} D.~J.,  2008, \mn@doi [Journal of Computational Physics]
  {10.1016/j.jcp.2008.08.011}, \href
  {http://adsabs.harvard.edu/abs/2008JCoPh.22710040P} {227, 10040}

\bibitem[\protect\citeauthoryear{{Price}}{{Price}}{2012a}]{Price2012a}
{Price} D.~J.,  2012a, \mn@doi [Journal of Computational Physics]
  {10.1016/j.jcp.2010.12.011}, \href
  {http://adsabs.harvard.edu/abs/2012JCoPh.231..759P} {231, 759}

\bibitem[\protect\citeauthoryear{{Price}}{{Price}}{2012b}]{Price2012b}
{Price} D.~J.,  2012b, \mn@doi [\mnras] {10.1111/j.1745-3933.2011.01187.x},
  \href {http://adsabs.harvard.edu/abs/2012MNRAS.420L..33P} {420, L33}

\bibitem[\protect\citeauthoryear{{Price} \& {Federrath}}{{Price} \&
  {Federrath}}{2010}]{Price2010}
{Price} D.~J.,  {Federrath} C.,  2010, \mn@doi [\mnras]
  {10.1111/j.1365-2966.2010.16810.x}, \href
  {http://adsabs.harvard.edu/abs/2010MNRAS.406.1659P} {406, 1659}

\bibitem[\protect\citeauthoryear{{Price} et~al.,}{{Price}
  et~al.}{2017}]{Price2017}
{Price} D.~J.,  et~al., 2017, preprint, \href
  {http://adsabs.harvard.edu/abs/2017arXiv170203930P} {} (\mn@eprint {arXiv}
  {1702.03930})

\bibitem[\protect\citeauthoryear{{Read}, {Hayfield}  \& {Agertz}}{{Read}
  et~al.}{2010}]{Read2010}
{Read} J.~I.,  {Hayfield} T.,   {Agertz} O.,  2010, \mn@doi [\mnras]
  {10.1111/j.1365-2966.2010.16577.x}, \href
  {http://adsabs.harvard.edu/abs/2010MNRAS.405.1513R} {405, 1513}

\bibitem[\protect\citeauthoryear{{Ritchie} \& {Thomas}}{{Ritchie} \&
  {Thomas}}{2001}]{Ritchie2001}
{Ritchie} B.~W.,  {Thomas} P.~A.,  2001, \mn@doi [\mnras]
  {10.1046/j.1365-8711.2001.04268.x}, \href
  {http://adsabs.harvard.edu/abs/2001MNRAS.323..743R} {323, 743}

\bibitem[\protect\citeauthoryear{{Rosswog}}{{Rosswog}}{2015}]{Rosswog2015}
{Rosswog} S.,  2015, \mn@doi [\mnras] {10.1093/mnras/stv225}, \href
  {http://adsabs.harvard.edu/abs/2015MNRAS.448.3628R} {448, 3628}

\bibitem[\protect\citeauthoryear{{Saitoh} \& {Makino}}{{Saitoh} \&
  {Makino}}{2009}]{Saitoh2009}
{Saitoh} T.~R.,  {Makino} J.,  2009, \mn@doi [\apjl]
  {10.1088/0004-637X/697/2/L99}, \href
  {http://adsabs.harvard.edu/abs/2009ApJ...697L..99S} {697, L99}

\bibitem[\protect\citeauthoryear{{Saitoh} \& {Makino}}{{Saitoh} \&
  {Makino}}{2013}]{Saitoh2013}
{Saitoh} T.~R.,  {Makino} J.,  2013, \mn@doi [\apj]
  {10.1088/0004-637X/768/1/44}, \href
  {http://adsabs.harvard.edu/abs/2013ApJ...768...44S} {768, 44}

\bibitem[\protect\citeauthoryear{{Schmidt}, {Niemeyer}  \&
  {Hillebrandt}}{{Schmidt} et~al.}{2006}]{Schmidt2006}
{Schmidt} W.,  {Niemeyer} J.~C.,   {Hillebrandt} W.,  2006, \mn@doi [\aap]
  {10.1051/0004-6361:20053617}, \href
  {http://adsabs.harvard.edu/abs/2006A%26A...450..265S} {450, 265}

\bibitem[\protect\citeauthoryear{{Sedov}}{{Sedov}}{1959}]{Sedov1959}
{Sedov} L.~I.,  1959, {Similarity and Dimensional Methods in Mechanics}

\bibitem[\protect\citeauthoryear{{Shen}, {Wadsley}  \& {Stinson}}{{Shen}
  et~al.}{2010}]{Shen2010}
{Shen} S.,  {Wadsley} J.,   {Stinson} G.,  2010, \mn@doi [\mnras]
  {10.1111/j.1365-2966.2010.17047.x}, \href
  {http://adsabs.harvard.edu/abs/2010MNRAS.407.1581S} {407, 1581}

\bibitem[\protect\citeauthoryear{{Sijacki}, {Vogelsberger}, {Kere{\v s}},
  {Springel}  \& {Hernquist}}{{Sijacki} et~al.}{2012}]{Sijacki2012}
{Sijacki} D.,  {Vogelsberger} M.,  {Kere{\v s}} D.,  {Springel} V.,
  {Hernquist} L.,  2012, \mn@doi [\mnras] {10.1111/j.1365-2966.2012.21466.x},
  \href {http://adsabs.harvard.edu/abs/2012MNRAS.424.2999S} {424, 2999}

\bibitem[\protect\citeauthoryear{Sod}{Sod}{1978}]{Sod1978}
Sod G.~A.,  1978, \mn@doi [Journal of Computational Physics]
  {http://dx.doi.org/10.1016/0021-9991(78)90023-2}, 27, 1

\bibitem[\protect\citeauthoryear{{Springel}}{{Springel}}{2005}]{Springel2005b}
{Springel} V.,  2005, \mn@doi [\mnras] {10.1111/j.1365-2966.2005.09655.x},
  \href {http://adsabs.harvard.edu/abs/2005MNRAS.364.1105S} {364, 1105}

\bibitem[\protect\citeauthoryear{{Springel}}{{Springel}}{2010a}]{Springel2010b}
{Springel} V.,  2010a, \mn@doi [\araa] {10.1146/annurev-astro-081309-130914},
  \href {http://adsabs.harvard.edu/abs/2010ARA%26A..48..391S} {48, 391}

\bibitem[\protect\citeauthoryear{{Springel}}{{Springel}}{2010b}]{Springel2010a}
{Springel} V.,  2010b, \mn@doi [\mnras] {10.1111/j.1365-2966.2009.15715.x},
  \href {http://adsabs.harvard.edu/abs/2010MNRAS.401..791S} {401, 791}

\bibitem[\protect\citeauthoryear{{Springel} \& {Hernquist}}{{Springel} \&
  {Hernquist}}{2002}]{Springel2002}
{Springel} V.,  {Hernquist} L.,  2002, \mn@doi [\mnras]
  {10.1046/j.1365-8711.2002.05445.x}, \href
  {http://adsabs.harvard.edu/abs/2002MNRAS.333..649S} {333, 649}

\bibitem[\protect\citeauthoryear{{Springel}, {Di Matteo}  \&
  {Hernquist}}{{Springel} et~al.}{2005}]{Springel2005a}
{Springel} V.,  {Di Matteo} T.,   {Hernquist} L.,  2005, \mn@doi [\mnras]
  {10.1111/j.1365-2966.2005.09238.x}, \href
  {http://adsabs.harvard.edu/abs/2005MNRAS.361..776S} {361, 776}

\bibitem[\protect\citeauthoryear{{Stadel}}{{Stadel}}{2001}]{Stadel2001}
{Stadel} J.~G.,  2001, PhD thesis, UNIVERSITY OF WASHINGTON

\bibitem[\protect\citeauthoryear{{Steinmetz} \& {Mueller}}{{Steinmetz} \&
  {Mueller}}{1993}]{Steinmetz1993}
{Steinmetz} M.,  {Mueller} E.,  1993, \aap, \href
  {http://adsabs.harvard.edu/abs/1993A%26A...268..391S} {268, 391}

\bibitem[\protect\citeauthoryear{{Tasker}, {Brunino}, {Mitchell}, {Michielsen},
  {Hopton}, {Pearce}, {Bryan}  \& {Theuns}}{{Tasker} et~al.}{2008}]{Tasker2008}
{Tasker} E.~J.,  {Brunino} R.,  {Mitchell} N.~L.,  {Michielsen} D.,  {Hopton}
  S.,  {Pearce} F.~R.,  {Bryan} G.~L.,   {Theuns} T.,  2008, \mn@doi [\mnras]
  {10.1111/j.1365-2966.2008.13836.x}, \href
  {http://adsabs.harvard.edu/abs/2008MNRAS.390.1267T} {390, 1267}

\bibitem[\protect\citeauthoryear{{Taylor}}{{Taylor}}{1950}]{Taylor1950}
{Taylor} G.,  1950, \mn@doi [Proceedings of the Royal Society of London Series
  A] {10.1098/rspa.1950.0049}, \href
  {http://adsabs.harvard.edu/abs/1950RSPSA.201..159T} {201, 159}

\bibitem[\protect\citeauthoryear{{Teyssier}}{{Teyssier}}{2002}]{Teyssier2002}
{Teyssier} R.,  2002, \mn@doi [\aap] {10.1051/0004-6361:20011817}, \href
  {http://adsabs.harvard.edu/abs/2002A%26A...385..337T} {385, 337}

\bibitem[\protect\citeauthoryear{{Theuns}, {Chalk}, {Schaller}  \&
  {Gonnet}}{{Theuns} et~al.}{2015}]{Theuns2016}
{Theuns} T.,  {Chalk} A.,  {Schaller} M.,   {Gonnet} P.,  2015, preprint, \href
  {http://adsabs.harvard.edu/abs/2015arXiv150800115T} {} (\mn@eprint {arXiv}
  {1508.00115})

\bibitem[\protect\citeauthoryear{{Valdarnini}}{{Valdarnini}}{2016}]{Valdarnini2016}
{Valdarnini} R.,  2016, preprint, \href
  {http://adsabs.harvard.edu/abs/2016arXiv160808361V} {} (\mn@eprint {arXiv}
  {1608.08361})

\bibitem[\protect\citeauthoryear{{Wadsley}, {Stadel}  \& {Quinn}}{{Wadsley}
  et~al.}{2004}]{Wadsley2004}
{Wadsley} J.~W.,  {Stadel} J.,   {Quinn} T.,  2004, New Astronomy, \href
  {http://adsabs.harvard.edu/cgi-bin/nph-bib_query?bibcode=2004NewA....9..137W&amp;db_key=AST}
  {9, 137}

\bibitem[\protect\citeauthoryear{{Wadsley}, {Veeravalli}  \&
  {Couchman}}{{Wadsley} et~al.}{2008}]{Wadsley2008}
{Wadsley} J.~W.,  {Veeravalli} G.,   {Couchman} H.~M.~P.,  2008, \mn@doi
  [\mnras] {10.1111/j.1365-2966.2008.13260.x}, \href
  {http://adsabs.harvard.edu/abs/2008MNRAS.387..427W} {387, 427}

\bibitem[\protect\citeauthoryear{Wendland}{Wendland}{1995}]{Wendland1995}
Wendland H.,  1995, \mn@doi [Advances in Computational Mathematics]
  {10.1007/BF02123482}, 4, 389

\bibitem[\protect\citeauthoryear{Woodward \& Colella}{Woodward \&
  Colella}{1984}]{Woodward1984}
Woodward P.,  Colella P.,  1984, \mn@doi [Journal of Computational Physics]
  {http://dx.doi.org/10.1016/0021-9991(84)90142-6}, 54, 115

\makeatother
\end{thebibliography}

%%%%%%%%%%%%%%%%%%%%%%%%%%%%%%%%%%%%%%%%%%%%%%%%%%

\bsp	
\label{lastpage}
\end{document}